\documentclass[12pt]{article}
\usepackage{amssymb}
\usepackage{rotating}
\usepackage{amsmath}
\usepackage{graphicx}
\usepackage{xcolor}
\usepackage{multirow}
\usepackage{array}
\usepackage{booktabs}
\usepackage{longtable}
\usepackage{arydshln}
\usepackage{etoolbox}
\makeatletter
\newcommand{\CXXIsetcolroom}{%
  \ifdim\pagegoal<\maxdimen
    \global\@colroom\pagegoal
  \else
    \global\@colroom\@colht
  \fi}
\patchcmd{\LT@start}
  {\global\@colroom\@colht}
  {\CXXIsetcolroom}
  {}
  {\errmessage{LT start patch failed}}
\patchcmd{\LT@output}
  {\global\@colroom\@colht}
  {\CXXIsetcolroom}
  {}
  {\errmessage{LT output patch failed}}
\makeatother
\usepackage{mathrsfs}
\usepackage{bm}
\usepackage[labelfont=bf]{caption}
\usepackage[colorlinks=true,linkcolor=blue,citecolor=blue,urlcolor=blue]{hyperref}
\usepackage{epstopdf}
\usepackage{seqsplit}
\usepackage{subfig}
\allowdisplaybreaks[4]

\graphicspath{{p/}}

\renewcommand{\theequation}{\arabic{section}.\arabic{equation}}

\renewcommand{\theHtable}{\theHsection.\arabic{table}}

\makeatletter
\AtBeginDocument{%
  \patchcmd{\LT@start}
    {\copy\ifvoid\LT@firsthead}
    {\Hy@raisedlink{\hyper@anchorstart{\@currentHref}\hyper@anchorend}%
     \copy\ifvoid\LT@firsthead}
    {}
    {\PackageError{C21}{Longtable hyperlink patch failed}{Check the longtable and arydshln versions.}}%
}
\makeatother

\begin{document}

%************************** Text Begins here ******************************

\newtheorem{Theorem}{Theorem}
\newtheorem{Definition}{Definition}
\newtheorem{Proposition}{Proposition}
\newtheorem{Lemma}{Lemma}
\newtheorem{Corollary}{Corollary}
\newcommand{\proof}[1]{{\bf Proof. }
        #1\begin{flushright}$\Box$\end{flushright}}

\baselineskip=20pt

%%%%%%%%%%%%%%%%%%%%%%%%%%%%%%%%%%%%%%%%%%%%%%%%%%%%%%%%%%%%
%                                                          %
%  Title page                                              %
%                                                          %
%%%%%%%%%%%%%%%%%%%%%%%%%%%%%%%%%%%%%%%%%%%%%%%%%%%%%%%%%%%%
\newfont{\elevenmib}{cmmib10 scaled\magstep1}
\newcommand{\preprint}{
   \begin{flushleft}
     %\elevenmib Yukawa\, Institute\, Kyoto\\
   \end{flushleft}\vspace{-1.3cm}
   \begin{flushright}\normalsize
  % \sf  YITP-03-53\\
   %  {\tt hep-th/yymmnnn} \\ November 2005
   \end{flushright}}
\newcommand{\Title}[1]{{\baselineskip=26pt
   \begin{center} \Large \bf #1 \\ \ \\ \end{center}}}

\newcommand{\Author}{%
	\begin{center}
		\large\bfseries
		Zhouzheng Ji${}^{a}$,
		Guang-Liang Li${}^{b,f}$,
		Yi Qiao${}^{a,e,f}$\footnote{Corresponding author: qiaoyi\_joy@foxmail.com},
		Junpeng Cao${}^{c,d,e}$\footnote{Corresponding author: junpengcao@iphy.ac.cn}
		and Wen-Li Yang${}^{a,e,f}$\footnote{Corresponding author: wlyang@nwu.edu.cn}
	\end{center}%
}

\newcommand{\Address}{%
	\begin{center}
		${}^{a}$ Institute of Modern Physics, Northwest University, Xi'an 710127, China\\
		${}^{b}$ Ministry of Education Key Laboratory for Nonequilibrium Synthesis and Modulation of Condensed Matter, School of Physics, Xi'an Jiaotong University, Xi'an 710049, China\\
		${}^{c}$ Beijing National Laboratory for Condensed Matter Physics, Institute of Physics, Chinese Academy of Sciences, Beijing 100190, China\\
		${}^{d}$ School of Physical Sciences, University of Chinese Academy of Sciences, Beijing 100049, China\\
		${}^{e}$ Peng Huanwu Center for Fundamental Theory, Xi'an 710127, China\\
		${}^{f}$ Shaanxi Key Laboratory for Theoretical Physics Frontiers, Xi'an 710127, China
	\end{center}%
}

\Title{Ground state and excitations of the \(C_2^{(1)}\) spin chain with non-diagonal boundaries}

\Author
\Address

\vspace{1cm}

\begin{abstract}
Understanding boundary effects in higher-rank integrable quantum spin chains requires connecting exact spectral relations with the root configurations underlying their energies and excitations.
With non-diagonal boundaries, determining these configurations remains challenging even when an exact spectral solution is available, hindering the direct extraction of thermodynamic properties.
We address this problem for the rational $C_2^{(1)}$ chain by combining DMRG in the tensor-network framework with the quantum inverse scattering method.
The numerically determined zeros of the fundamental and fused transfer-matrix eigenvalues reveal three bulk-string distributions, labeled by $x$, $y$, and $w$, together with boundary strings and additional roots.
Substituting these configurations into the operator identities relating the two transfer matrices yields the ground-state energy and the surface energy in the thermodynamic limit under non-diagonal boundary conditions.
The analytical surface energy agrees with the numerical results up to $O(\tfrac1N)$ corrections.
From the regularity conditions of the operator identities, we derive a boundary-string selection rule and numerically identify allowed configurations beyond the ground state.
For excitations preserving the boundary-string configuration, we identify three elementary hole excitations, $X$, $Y$, and $W$, in the $x$-, $y$-, and $w$-string distributions, respectively, and derive their energies.
Based on the symmetry of the zero roots, we also present a bulk-excitation selection rule governing the allowed combinations of these elementary excitations in physical eigenstates.
The framework developed here can therefore be used to systematically analyze boundary effects, bulk excitations, and related physical properties in other high-rank integrable quantum models with non-diagonal boundary conditions.

\vspace{0.5truecm}
\noindent {\it Keywords}: quantum spin chains; integrability; DMRG; exact solution
\end{abstract}
\newpage
\section{Introduction}
\label{intro}
\setcounter{equation}{0}

Quantum integrable systems provide exact results for the spectra and thermodynamic properties of interacting many-body systems \cite{baxter,korepin}.
Within the quantum inverse scattering method, the Yang--Baxter equation generates commuting transfer matrices, from which integrable Hamiltonians can be constructed.
Bethe ansatz and functional relations allow their spectral properties to be studied and their ground-state properties and elementary excitations to be determined.
These methods connect microscopic interactions with macroscopic properties, providing a framework for understanding collective behavior in one dimension.

The rational $C_2^{(1)}$ chain is a rank-two integrable model with a four-dimensional local space, admitting both spin-$3/2$ and spin-orbital representations \cite{li}.
Although it shares the same local dimension with the $\mathrm{SU}(4)$ chain, their different algebraic structures lead to distinct excitation spectra.
For the periodic $C_2^{(1)}$ chain, Bethe ansatz analysis identifies three coupled root distributions supporting two spinon branches and a convolution-type composite branch \cite{martins}.
This model therefore provides a relatively simple setting for studying coupled excitations in higher-rank integrable systems and for comparing the physical consequences of different algebraic structures at the same local dimension.
Related developments include integrable two-leg spin ladders with $sp(4)$ symmetry \cite{batchelor} and the determination of the partition function per site for periodic $Sp(4)$ and $Sp(2n)$ vertex models in the thermodynamic limit through fusion relations \cite{ribeiro2022,ribeiro2023}.
Extending these studies to the open chain allows us to investigate how boundary interactions affect the energy, root configurations, and excitation structure.

Integrable boundary interactions can be introduced through reflection matrices satisfying the reflection equations \cite{sklyanin,devega}.
For generic off-diagonal boundaries, however, the conventional algebraic Bethe ansatz generally lacks a simple reference state \cite{odba}.
The off-diagonal Bethe ansatz provides a way to treat integrable models without $U(1)$ symmetry \cite{cao2013} and, combined with fusion, has yielded exact solutions of the open $C_2^{(1)}$ chain \cite{li2019,li}.
Nevertheless, the resulting Bethe ansatz equations contain inhomogeneous terms, and the Bethe-root configurations of the relevant physical states are complicated, making the direct extraction of thermodynamic properties difficult.
Identifying the ground-state and excited-state root configurations and using them to derive thermodynamic properties and excitation spectra thus remain necessary steps beyond obtaining the exact spectral relations.

Fusion connects transfer matrices in different auxiliary representations and yields additional relations among their eigenvalues \cite{karowski,kulish,mezincescu}.
In higher-rank models, these coupled relations reflect the underlying representation structure, as illustrated by the fusion construction for $\mathrm{SU}(n)$ chains \cite{lu2021}.
For the open $C_2^{(1)}$ chain, the fundamental transfer matrix and its five-dimensional fused counterpart satisfy closed operator identities \cite{li}.
Parametrizing their eigenvalues by their zeros recasts these identities as coupled relations between two root sets.
Unlike the Bethe-root representation, this approach describes spectral properties directly in terms of the zeros of transfer-matrix eigenvalues, providing an alternative route to thermodynamic quantities \cite{qiao2,qiao1}.
It has been used to obtain surface and excitation energies in the XXX, XXZ, and XYZ chains with off-diagonal boundaries \cite{dong2023,qiao3,xin2024}.
For the open $C_2^{(1)}$ chain, Ref.~\cite{li} also pointed out the need to parametrize the transfer-matrix eigenvalues by their zeros to study the thermodynamic limit.

However, obtaining clear root structures requires numerical zero-root distributions for sufficiently long chains, where directly solving the coupled zero-root equations is already difficult, let alone directly diagonalizing the transfer matrices.
The BA-DMRG method developed in Ref.~\cite{ji}, which was used to obtain zero-root configurations of the XXX chain, offers a practical approach to this problem.
It first uses the density-matrix renormalization group (DMRG) in the tensor-network framework to obtain matrix product state representations of Hamiltonian eigenstates \cite{schollwock2011}.
The corresponding transfer-matrix eigenvalues are then evaluated using matrix product operators, and their zeros are extracted without directly solving the coupled zero-root equations.
Here we apply this method to both the fundamental and fused transfer matrices of the open $C_2^{(1)}$ chain.
The resulting configurations reveal three bulk-string distributions, associated with the $x$-, $y$-, and $w$-strings, together with boundary strings and additional roots, providing the basis for our analytical treatment.

Substituting these root configurations into the operator identities gives three coupled root-density equations in the thermodynamic limit.
For the ground state, we solve the corresponding equations to obtain the root densities and energy under non-diagonal boundary conditions, and extract the surface energy by subtracting the extensive bulk contribution.
We further compare the analytical surface energy with the difference between the open-chain ground-state energy obtained by DMRG and the periodic-chain ground-state energy calculated from the Bethe ansatz equations.
The numerical results support the analytical expression, with deviations consistent with $O(1/N)$ finite-size corrections.

The operator identities also allow boundary-string configurations beyond those realized in the ground state.
Requiring regularity of the Fourier-transformed root densities determines the possible boundary-string positions and yields a selection rule for their admissible combinations.
We also calculate the excitation energies associated with changes in the boundary-string configuration.
For a given boundary-string configuration, its lowest-energy state is termed the base state, following Ref.~\cite{zhakenov2025}.
In numerical excited-state calculations, we identify additional boundary-string configurations that satisfy this selection rule.

For excitations preserving the boundary-string configuration, we derive the density responses and energies of three elementary hole excitations, $X$, $Y$, and $W$, associated with the $x$-, $y$-, and $w$-string distributions, respectively.
After accounting for the different normalization, the resulting elementary bulk excitation energies agree with those of the periodic chain \cite{martins}.
Thus, even when the boundaries are non-diagonal, the zero-root representation reveals an elementary bulk excitation structure closely related to that of the periodic chain.
The numbers and admissible combinations of these excitations are constrained by a bulk-excitation selection rule derived from changes in string occupation numbers and zero-root symmetry.
This rule is supported by the numerically examined excited-state configurations.
Changes in boundary-string content may also require accompanying bulk excitations to satisfy the fixed zero-root counts and zero-root symmetry.
Together, these results provide a zero-root description of boundary effects and bulk excitations in the open $C_2^{(1)}$ chain with non-diagonal boundaries.

The paper is organized as follows.
Section~\ref{sec:integrability} introduces the model, fusion construction, and operator identities.
Section~\ref{sec:ground-state} presents the ground-state root configurations and energy.
Section~\ref{sec:boundary-strings} determines the ground-state boundary strings and the selection rule for their admissible combinations, derives their excitation energies and establishes the uniqueness of the minimum-energy boundary-locked configuration.
Section~\ref{sec:excitations} studies the elementary bulk excitations and their selection rule.
Section~\ref{sec:discussion} summarizes the results.
Appendix~\ref{app:A} gives the fused reflection matrix, and Appendix~\ref{app:numerical-boundary-roots} compares the analytical and numerical boundary-string positions.
Appendices~\ref{app:W-center-locking} and \ref{app:zero-root-symmetry} derive the $W$-type root configuration from density regularity and establish the symmetry of the zero-root distributions, respectively.
Appendix~\ref{app:numerical-excitation-counts} lists the numerical excitation configurations and checks the bulk-excitation selection rule.

\section{Model and Integrable Structure}
\label{sec:integrability}
\setcounter{equation}{0}
\subsection{Hamiltonian and Transfer Matrix}
We follow the integrable construction and conventions of Ref.~\cite{li}.
The construction of the integrable \(C_2^{(1)}\) model starts from the following \(R\)-matrix \cite{martinsramos}:
\begin{equation}
\setlength{\arraycolsep}{3.5pt}
\renewcommand{\arraystretch}{0.95}
R_{12}(u)=
\left(
\begin{array}{cccc|cccc|cccc|cccc}
a &   &   &   &   &   &   &   &   &   &   &   &   &   &   &   \\
  & b &   &   & g &   &   &   &   &   &   &   &   &   &   &   \\
  &   & b &   &   &   &   &   & g &   &   &   &   &   &   &   \\
  &   &   & e &   &   & d &   &   & -d &   &   & c &   &   &   \\
\hline
  & g &   &   & b &   &   &   &   &   &   &   &   &   &   &   \\
  &   &   &   &   & a &   &   &   &   &   &   &   &   &   &   \\
  &   &   & d &   &   & e &   &   & c &   &   & -d &   &   &   \\
  &   &   &   &   &   &   & b &   &   &   &   &   & g &   &   \\
\hline
  &   & g &   &   &   &   &   & b &   &   &   &   &   &   &   \\
  &   &   & -d &   &   & c &   &   & e &   &   & d &   &   &   \\
  &   &   &   &   &   &   &   &   &   & a &   &   &   &   &   \\
  &   &   &   &   &   &   &   &   &   &   & b &   &   & g &   \\
\hline
  &   &   & c &   &   & -d &   &   & d &   &   & e &   &   &   \\
  &   &   &   &   &   &   & g &   &   &   &   &   & b &   &   \\
  &   &   &   &   &   &   &   &   &   &   & g &   &   & b &   \\
  &   &   &   &   &   &   &   &   &   &   &   &   &   &   & a
\end{array}
\right),
\label{eq:Rmatrix}
\end{equation}
which is defined in the tensor space \(V_1\otimes V_2\), with \(a(u)=(u+1)(u+3)\), \(b(u)=u(u+3)\), \(c(u)=2u+3\), \(d(u)=-u\), \(e(u)=u(u+2)\) and \(g(u)=u+3\).
The \(R\)-matrix~\eqref{eq:Rmatrix} has the following properties
\begin{align}
\text{regularity:}\quad 
& R_{12}(0)=\rho_1(0)^{\frac12}\mathcal P_{12}, \nonumber\\
\text{unitarity:}\quad 
& R_{12}(u)R_{21}(-u)=\rho_1(u), \nonumber\\
\text{crossing-unitarity:}\quad 
& R_{12}(u)^{t_1}R_{21}(-u-6)^{t_1}=\rho_1(u+3),
\end{align}
where \(\rho_1(u)=(u^2-1)(u^2-9)\), \(\mathcal P_{12}\) is the permutation operator with the elements \(\mathcal P_{kl}^{ij}=\delta_{il}\delta_{jk}\), and \(t_i\) denotes the transposition in the \(i\)-th space.
Moreover, \(R_{21}(u)=\mathcal P_{12}R_{12}(u)\mathcal P_{12}=R_{12}^{t_1t_2}(u)\).
The \(R\)-matrix satisfies the Yang--Baxter equation (YBE)
\begin{equation}
R_{12}(u-v)R_{13}(u)R_{23}(v)
=
R_{23}(v)R_{13}(u)R_{12}(u-v).
\end{equation}

Boundary reflections are described by reflection matrices \cite{devega,devega2,lisantos,malara}.
Here we adopt the simplified parametrization of Ref.~\cite{li}, for which the reflection matrix at one boundary takes the form
\begin{equation}
K_1^{-}(u)=1+
\left(
\begin{array}{cccc}
-c_1 & c_2 & c_1 & c_2 \\
c_2 & -c_1 & c_2 & c_1 \\
c_1 & c_2 & c_1 & -c_2 \\
c_2 & c_1 & -c_2 & c_1
\end{array}
\right)u,
\end{equation}
which satisfies the reflection equation (RE)
\begin{equation}
R_{12}(u-v)K_1^-(u)R_{21}(u+v)K_2^-(v)
=
K_2^-(v)R_{12}(u+v)K_1^-(u)R_{21}(u-v).
\end{equation}
The parametrization used in this work is a simplified form of the model in Ref.~\cite{li}; both $c_1$ and $c_2$ are treated as free parameters.

According to the quantum inverse scattering method, the boundary reflection at the other side is described by the dual reflection matrix $K_1^{+}(u)$, which can be obtained from the mapping
\begin{equation}
K_1^{+}(u)=K_1^{-}(-u-3)\big|_{\{c_1,c_2\}\to\{\tilde c_1,\tilde c_2\}},
\end{equation}
where $\{\tilde c_1,\tilde c_2\}$ are the right-boundary parameters.
The dual reflection equation for the boundary $K^+(u)$-matrix is given by
\begin{equation}
R_{12}(-u+v)K_1^+(u)R_{21}(-u-v-6)K_2^+(v)
=
K_2^+(v)R_{12}(-u-v-6)K_1^+(u)R_{21}(-u+v).
\end{equation}

With the $R$-matrix introduced above, the monodromy matrices are constructed as
\begin{align}
T_0(u)&=R_{0N}(u-\theta_N)R_{0,N-1}(u-\theta_{N-1})\cdots R_{01}(u-\theta_1),\\
\hat T_0(u)&=R_{10}(u+\theta_1)\cdots R_{N-1,0}(u+\theta_{N-1})R_{N0}(u+\theta_N),
\end{align}
where $\{\theta_j\mid j=1,\ldots,N\}$ are the inhomogeneous parameters.
The subscript $0$ labels the auxiliary space, while the indices $j=1,\ldots,N$ denote the quantum spaces.
As a consequence of the Yang--Baxter equation, the monodromy matrices obey
\begin{align}
R_{12}(u-v)T_1(u)T_2(v)&=T_2(v)T_1(u)R_{12}(u-v),\\
R_{21}(u-v)\hat T_1(u)\hat T_2(v)&=\hat T_2(v)\hat T_1(u)R_{21}(u-v).
\end{align}
Taking the boundary reflections into account, the transfer matrix is defined by \cite{sklyanin}
\begin{equation}
t(u)=\operatorname{tr}_0\!\left\{K_0^+(u)T_0(u)K_0^-(u)\hat T_0(u)\right\},
\label{eq:transfer-matrix}
\end{equation}
where $\operatorname{tr}_0$ denotes the trace over the auxiliary space.
Using the Yang--Baxter equation together with the reflection equation and the dual reflection equation, one obtains the commutativity relation
\begin{equation}
[t(u),t(v)]=0 .
\end{equation}
Thus, $t(u)$ generates a family of mutually commuting conserved quantities, and the Hamiltonian of the open spin chain is obtained from the transfer matrix as
\begin{align}
H
&=
\left.\frac{3}{2}\frac{\partial \ln t(u)}{\partial u}\right|_{u=0,\{\theta_j\}=0}-\frac{3\operatorname{tr}_{0} K_{0}^{+\prime}(0)}{2\operatorname{tr}_{0} K_{0}^{+}(0)}
\nonumber \\
&=
\sum_{k=1}^{N-1} H_{k,k+1}+\frac{3K_{1}^{-\prime}(0)}{2}+\frac{\operatorname{tr}_{0}\!\left\{K_{0}^{+}(0)H_{N0}\right\}}{\operatorname{tr}_{0}K_{0}^{+}(0)},
\label{eq:Hamiltonian-general}
\end{align}
where $H_{k,k+1}=\mathcal P_{k,k+1}R'_{k,k+1}(u)\big|_{u=0}$.
The model Hamiltonian in Eq.~\eqref{eq:Hamiltonian-general} can be expressed in terms of either spin-$3/2$ operators, which characterize the corresponding spin-exchange interactions, or two sets of Pauli matrices, which describe the associated spin-orbital couplings \cite{li,batchelor}.

Using the Pauli matrices $\vec{\sigma}$ in the spin degree of freedom and $\vec{\tau}$ in the orbital degree of freedom, the model Hamiltonian can be written as
\begin{equation}
\begin{aligned}
H=&\sum_{i=1}^{N-1}\left\{\frac12(\vec{\sigma}_i\!\cdot\!\vec{\sigma}_{i+1})
\left(1+\vec{\tau}_i\!\cdot\!\vec{\tau}_{i+1}+\tau_i^z\tau_{i+1}^z\right)
+\vec{\tau}_i\!\cdot\!\vec{\tau}_{i+1}-\frac12\tau_i^z\tau_{i+1}^z\right\} \\
&+\frac32\left[c_1(\sigma_1^x-\sigma_1^z)+c_2(\sigma_1^z+\sigma_1^x)\tau_1^x\right]
-\frac32\left[\tilde c_1(\sigma_N^x-\sigma_N^z)+\tilde c_2(\sigma_N^z+\sigma_N^x)\tau_N^x\right]+2N ,
\end{aligned}
\label{eq:Hamiltonian-pauli}
\end{equation}
where $\vec{\sigma}_j=(\sigma_j^x,\sigma_j^y,\sigma_j^z)$ and $\vec{\tau}_j=(\tau_j^x,\tau_j^y,\tau_j^z)$ are two independent sets of Pauli matrices acting on the local space at site $j$.

The $C_2\simeq\mathfrak{sp}(4)$ symmetry also makes the two-site representation content explicit.
In the spin-orbital basis, the four local states are
\begin{equation}
|1\rangle=|\uparrow,\uparrow\rangle,\qquad
|2\rangle=|\uparrow,\downarrow\rangle,\qquad
|3\rangle=|\downarrow,\uparrow\rangle,\qquad
|4\rangle=|\downarrow,\downarrow\rangle .
\end{equation}
The arrows label the independent $\sigma$ and $\tau$ degrees of freedom.
The tensor product of two neighbouring local spaces is
\begin{equation}
\mathcal H_i\otimes\mathcal H_{i+1}\simeq 4\otimes4 .
\end{equation}
The $16$-dimensional two-site space decomposes under the $C_2\simeq\mathfrak{sp}(4)$ symmetry as
\begin{equation}
4\otimes4=10\oplus5\oplus1 .
\end{equation}
The ten-dimensional component corresponds to the symmetric subspace under the exchange of the two four-state local spaces.
The remaining six-dimensional antisymmetric subspace would remain irreducible for an $SU(4)$-symmetric structure, whereas it further splits into a five-dimensional component and a singlet under $C_2\simeq\mathfrak{sp}(4)$,
\begin{equation}
6=5\oplus1 .
\end{equation}
The singlet is spanned by
\begin{equation}
|\Omega\rangle=
\frac{1}{2}
\left(
|14\rangle+|23\rangle-|32\rangle-|41\rangle
\right).
\end{equation}
In this representation, the local bulk Hamiltonian can be written in the compact form
\begin{equation}
H_{i,i+1}
=1+3\mathcal P_{i,i+1}+4|\Omega\rangle\langle\Omega| .
\end{equation}

\subsection{Fusion Relations and Operator Identities}
In this subsection, we review the fused transfer matrix and the closed set of operator identities obtained in Ref.~\cite{li} using the fusion method \cite{karowski,kulish,mezincescu,zhou}.

The fused monodromy matrices are constructed from the fused $\bar R$-matrices $\bar R_{\bar 0 j}(u)$, which are defined on the tensor product of the fused auxiliary space $\bar V_0$ and the quantum space $V_j$.
For instance, $\bar R_{\bar 1 3}(u)$ can be written explicitly as
\begin{equation}
\bar R_{\bar 1 3}(u)
=
\left[
\left(u+\frac{3}{2}\right)
\tilde\rho_0\left(u+\frac{1}{2}\right)
\right]^{-1}
P^{(5)}_{12}
R_{23}\left(u+\frac{1}{2}\right)
R_{13}\left(u-\frac{1}{2}\right)
P^{(5)}_{12}.
\label{eq:fused-Rbar}
\end{equation}
Here $\tilde\rho_0(u)=(u-1)(u+3)$, and $P^{(5)}_{12}$~\cite{li} projects the two fundamental auxiliary spaces onto the five-dimensional component.
In the resulting basis, the fused $\bar R$-matrix reads
\begin{equation}
\setlength{\arraycolsep}{1.5pt}
\renewcommand{\arraystretch}{1.1}
\bar{R}_{\bar{1}3}(u)=
\left(
\begin{array}{cccc|cccc|cccc|cccc|cccc}
a_1 &     &     &     &     &     &     &     &     &     &     &     &      &     &     &     &      &     &     &     \\
    & a_1 &     &     &     &     &     &     &     &     &     &     &      &     &     &     &      &     &     &     \\
    &     & e_1 &     &     & d_1 &     &     & g_1 &     &     &     &      &     &     &     &      &     &     &     \\
    &     &     & e_1 &     &     &     &     &     & g_1 &     &     & -d_1 &     &     &     &      &     &     &     \\
\hline
    &     &     &     & a_1 &     &     &     &     &     &     &     &      &     &     &     &      &     &     &     \\
    &     & d_1 &     &     & e_1 &     &     & -g_1&     &     &     &      &     &     &     &      &     &     &     \\
    &     &     &     &     &     & a_1 &     &     &     &     &     &      &     &     &     &      &     &     &     \\
    &     &     &     &     &     &     & e_1 &     &     & g_1 &     &      &     &     &     & -d_1 &     &     &     \\
\hline
    &     & g_1 &     &     & -g_1&     &     & b_1 &     &     &     &      &     &     &     &      &     &     &     \\
    &     &     & g_1 &     &     &     &     &     & b_1 &     &     & g_1  &     &     &     &      &     &     &     \\
    &     &     &     &     &     &     & g_1 &     &     & b_1 &     &      &     &     &     & g_1  &     &     &     \\
    &     &     &     &     &     &     &     &     &     &     & b_1 &      &     & -g_1&     &      & g_1 &     &     \\
\hline
    &     &     & -d_1&     &     &     &     &     & g_1 &     &     & e_1  &     &     &     &      &     &     &     \\
    &     &     &     &     &     &     &     &     &     &     &     &      & a_1 &     &     &      &     &     &     \\
    &     &     &     &     &     &     &     &     &     &     & -g_1&      &     & e_1 &     &      & d_1 &     &     \\
    &     &     &     &     &     &     &     &     &     &     &     &      &     &     & a_1 &      &     &     &     \\
\hline
    &     &     &     &     &     &     & -d_1&     &     & g_1 &     &      &     &     &     & e_1  &     &     &     \\
    &     &     &     &     &     &     &     &     &     &     & g_1 &      &     & d_1 &     &      & e_1 &     &     \\
    &     &     &     &     &     &     &     &     &     &     &     &      &     &     &     &      &     & a_1 &     \\
    &     &     &     &     &     &     &     &     &     &     &     &      &     &     &     &      &     &     & a_1
\end{array}
\right),
\end{equation}
where $a_1=u+\frac{5}{2}$, $b_1=u+\frac{3}{2}$, $e_1=u+\frac{1}{2}$, $d_1=2$, and $g_1=\sqrt{2}$.
Thus, all matrix elements are polynomials of degree at most one in $u$.
Moreover, the fused $\bar R$-matrix satisfies the Yang--Baxter equation
\begin{equation}
\bar R_{\bar 1 2}(u-v)\bar R_{\bar 1 3}(u)R_{23}(v)
=
R_{23}(v)\bar R_{\bar 1 3}(u)\bar R_{\bar 1 2}(u-v).
\end{equation}

The fused reflection matrices in the fused auxiliary space are defined as
\begin{align}
\bar K_{\bar 1}^{-}(u)
&=
-\left[(2u-1)(2u+3)\right]^{-1}
P_{21}^{(5)}
K_1^{-}\left(u+\frac{1}{2}\right)
R_{21}(2u)
K_2^{-}\left(u-\frac{1}{2}\right)
P_{12}^{(5)}
\nonumber\\
&=
\begin{pmatrix}
\bar k_{11} & \bar k_{12} & \bar k_{13} & \bar k_{14} & \bar k_{15}\\
\bar k_{21} & \bar k_{22} & \bar k_{23} & \bar k_{24} & \bar k_{25}\\
\bar k_{31} & \bar k_{32} & \bar k_{33} & \bar k_{34} & \bar k_{35}\\
\bar k_{41} & \bar k_{42} & \bar k_{43} & \bar k_{44} & \bar k_{45}\\
\bar k_{51} & \bar k_{52} & \bar k_{53} & \bar k_{54} & \bar k_{55}
\end{pmatrix},
\label{eq:Kfused}
\end{align}
\begin{equation}
\begin{aligned}
\bar K_{\bar 1}^{+}(u)
&=
-\left[(2u+3)(2u+7)\right]^{-1}
P_{12}^{(5)}
K_2^{+}\left(u-\frac{1}{2}\right)
R_{12}(-2u-6)
K_1^{+}\left(u+\frac{1}{2}\right)
P_{21}^{(5)}\\
&=
\left.
\bar K_{\bar 1}^{-}(-u-3)
\right|_{\{c_1,c_2\}\to\{\tilde c_1,\tilde c_2\}}.
\end{aligned}
\end{equation}
The explicit forms of the matrix elements are listed in Appendix~\ref{app:A}.
It should be noted that $\bar K^{\pm}(u)$ are both $5\times5$ matrices, whose elements are polynomials in $u$ of degree at most two.
The fused $\bar R$-matrix and the fused reflection matrices $\bar K^{\pm}(u)$ satisfy the reflection equations
\begin{equation}
\bar R_{\bar1 2}(u-v)\bar K_{\bar1}^{-}(u)
\bar R_{2\bar1}(u+v)K_2^{-}(v)
=
K_2^{-}(v)\bar R_{\bar1 2}(u+v)
\bar K_{\bar1}^{-}(u)\bar R_{2\bar1}(u-v).
\label{eq:fused-RE-minus}
\end{equation}
\begin{equation}
\begin{aligned}
&\bar R_{\bar1 2}(-u+v)\bar K_{\bar1}^{+}(u)
\bar R_{2\bar1}(-u-v-6)K_2^{+}(v)\\
&\qquad =
K_2^{+}(v)\bar R_{\bar1 2}(-u-v-6)
\bar K_{\bar1}^{+}(u)\bar R_{2\bar1}(-u+v).
\end{aligned}
\label{eq:fused-RE-plus}
\end{equation}
We define the fused transfer matrix $\bar t(u)$ as
\begin{equation}
\bar t(u)=\operatorname{tr}_{\bar0}
\left\{
\bar K_{\bar0}^{+}(u)\bar T_{\bar0}(u)\bar K_{\bar0}^{-}(u)\hat{\bar T}_{\bar0}(u)
\right\}.
\label{eq:fused-transfer-matrix}
\end{equation}
The transfer matrix $t(u)$ and its fused one $\bar{t}(u)$ form a mutually commuting family:
\begin{equation}
[t(u),\bar t(v)]=0 .
\label{eq:t-tbar-commutativity}
\end{equation}
Therefore, $t(u)$ and $\bar t(v)$ share a common set of eigenstates.

Let $\Lambda(u)$ and $\bar\Lambda(u)$ be the eigenvalues of $t(u)$ and $\bar t(u)$, respectively.
In addition, these two transfer matrices possess the crossing symmetry
\begin{equation}
t(u)=t(-u-3),\qquad \bar t(u)=\bar t(-u-3).
\label{eq:crossing-symmetry}
\end{equation}
Therefore, their eigenvalues $\Lambda(u)$ and $\bar\Lambda(u)$ satisfy
\begin{equation}
\Lambda(u)=\Lambda(-u-3),\qquad 
\bar\Lambda(u)=\bar\Lambda(-u-3).
\label{eq:eigenvalue-crossing}
\end{equation}
After eliminating the redundant constraints implied by crossing symmetry, the following three independent functional relations are retained:
\begin{equation}
\begin{aligned}
\Lambda(\theta_j)\Lambda(\theta_j-3)
&=
\frac{1}{2^4}
\frac{(\theta_j-1)(\theta_j-3)(\theta_j+1)(\theta_j+3)}
{\left(\theta_j-\frac32\right)\left(\theta_j-\frac12\right)
 \left(\theta_j+\frac32\right)\left(\theta_j+\frac12\right)}
h(\theta_j)\tilde h(\theta_j) \\
&\quad\times
\prod_{i=1}^{N}
a(\theta_j-\theta_i)a(\theta_j+\theta_i)
e(\theta_j-\theta_i-3)e(\theta_j+\theta_i-3),
\qquad j=1,\ldots,N .
\end{aligned}
\label{eq:I}
\end{equation}
\begin{equation}
\begin{aligned}
\Lambda(\theta_j)\Lambda(\theta_j-1)
&=
\frac{(\theta_j-1)(\theta_j+1)^2(\theta_j+3)}
{\left(\theta_j-\frac12\right)\left(\theta_j+\frac12\right)
 \left(\theta_j+\frac32\right)\left(\theta_j+\frac52\right)} \\
&\quad\times
\prod_{i=1}^{N}
(\theta_j-\theta_i+1)(\theta_j+\theta_i+1)
\tilde\rho_0(\theta_j-\theta_i)\tilde\rho_0(\theta_j+\theta_i) \\
&\quad\times
\bar\Lambda\left(\theta_j-\frac12\right),
\qquad j=1,\ldots,N .
\end{aligned}
\label{eq:II}
\end{equation}
\begin{equation}
\begin{aligned}
\Lambda(\theta_j-3)\Lambda(\theta_j-2)
&=
\frac{(-\theta_j-1)(-\theta_j+1)^2(-\theta_j+3)}
{\left(-\theta_j-\frac12\right)\left(-\theta_j+\frac12\right)
 \left(-\theta_j+\frac32\right)\left(-\theta_j+\frac52\right)} \\
&\quad\times
\prod_{i=1}^{N}
(-\theta_j-\theta_i+1)(-\theta_j+\theta_i+1)
\tilde\rho_0(-\theta_j-\theta_i)\tilde\rho_0(-\theta_j+\theta_i) \\
&\quad\times
\bar\Lambda\left(\theta_j-\frac52\right),
\qquad j=1,\ldots,N .
\end{aligned}
\label{eq:III}
\end{equation}
The functions $h(u)$ and $\tilde h(u)$ are defined as
\begin{align}
h(u)&=4(Cu^2-1), \qquad C=2(c_1^2+c_2^2), \nonumber\\
\tilde h(u)&=4(\tilde C u^2-1), \qquad 
\tilde C=2(\tilde c_1^2+\tilde c_2^2).
\label{eq:h(u)}
\end{align}

From Eqs.~\eqref{eq:transfer-matrix} and \eqref{eq:fused-transfer-matrix}, $t(u)$ and $\bar t(u)$ are polynomials in $u$ of degrees $4N+2$ and $2N+4$, respectively, for generic boundary parameters.
The crossing symmetry~\eqref{eq:eigenvalue-crossing} pairs their eigenvalue zeros about $u=-3/2$.
We therefore introduce the zero-root parameters $\{z_j\}_{j=1}^{2N+1}$ and $\{\bar z_j\}_{j=1}^{N+2}$ through
\begin{equation}
\Lambda(u)
=-A\prod_{j=1}^{2N+1}
\left(u+i z_j+\frac{3}{2}\right)
\left(u-i z_j+\frac{3}{2}\right),
\label{eq:Lambda-root}
\end{equation}
where
\begin{equation}
A=8(c_1\tilde c_1+c_2\tilde c_2),
\label{eq:A(u)}
\end{equation}
and
\begin{equation}
\bar\Lambda(u)
=\left(\frac{A^2}{4}+C\tilde C\right)
\prod_{j=1}^{N+2}
\left(u+i\bar z_j+\frac{3}{2}\right)
\left(u-i\bar z_j+\frac{3}{2}\right).
\label{eq:Lambda-bar-root}
\end{equation}
Hereafter, we refer to $\{\pm z_j\}$ and $\{\pm\bar z_j\}$ as the zero roots; the corresponding zeros in the spectral $u$-plane are $u=-3/2\pm i z_j$ and $u=-3/2\pm i\bar z_j$, respectively.

\section{Ground-State Root Structure and Energy}
\label{sec:ground-state}
\setcounter{equation}{0}

In this section, we determine the ground-state zero-root configurations and combine them with the operator identities in Eqs.~\eqref{eq:I}--\eqref{eq:III} to derive the root densities and ground-state energy in the thermodynamic limit.

\subsection{Ground-State Root Configurations}

We first determine the ground-state zero-root distributions of $\Lambda(u)$ and $\bar\Lambda(u)$ for the homogeneous chain using the BA-DMRG method of Ref.~\cite{ji}.

The DMRG calculations used a maximum bond dimension of 300 and a truncation cutoff of $10^{-10}$.
At least 10 sweeps were performed, with convergence determined by a relative energy difference below $10^{-10}$ between consecutive sweeps.

By examining the zero-root distributions  for chain lengths $N\equiv0\pmod{4}$ under a wide range of boundary parameters $c_1$, $c_2$, $\tilde c_1$, and $\tilde c_2$, we find that the ground-state root structures can be effectively organized by the boundary combinations $C$, $\tilde C$, and $A$ introduced above.

This parametrization is consistent with the boundary dependence appearing in the functional relations.
As shown in Eq.~\eqref{eq:h(u)}, the combinations $C$ and $\tilde C$ enter through the two boundary factors $h(u)$ and $\tilde h(u)$.
The corresponding characteristic spectral positions are given by $1/\sqrt C$ and $1/\sqrt{\tilde C}$.
Based on this observation and on the numerical zero-root distributions, we introduce
\begin{equation}
s_{\rm L}=\frac{1}{\sqrt C},\qquad 
s_{\rm R}=\frac{1}{\sqrt{\tilde C}},
\label{eq:boundary-spectral-variables}
\end{equation}
and use them as the two coordinates for presenting the phase diagrams.

The remaining combination $A$ in Eq.~\eqref{eq:A(u)} contains the relative-orientation information between the two boundary parameter pairs.
After the boundary scales are fixed by $C$ and $\tilde C$, the two pairs $(c_1,c_2)$ and $(\tilde c_1,\tilde c_2)$ still carry an angular degree of freedom in the two-component boundary-parameter space.
As shown in Eq.~\eqref{eq:A(u)}, this angular dependence is encoded in the relative-orientation factor involving $A$.
Numerically, we find that the zero-root structures should be classified separately in the two sectors $A>0$ and $A<0$.

\begin{figure}[htbp]
\centering
\includegraphics[width=0.5\textwidth]{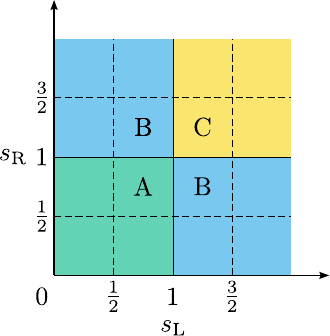}
\caption{Ground-state phase diagram of the zero-root configurations $\{\pm z_j\}$ of $\Lambda(u)$ for $A>0$ in the $(s_{\rm L},s_{\rm R})$ plane.
The horizontal axis is $s_{\rm L}=1/\sqrt C$, and the vertical axis is $s_{\rm R}=1/\sqrt{\tilde C}$.}
\label{fig:Lambda+}
\end{figure}

\begin{figure}[htbp]
\centering
\includegraphics[width=0.5\textwidth]{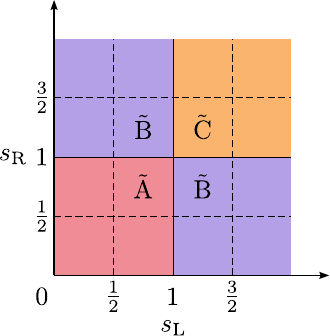}
\caption{Ground-state phase diagram of the zero-root configurations $\{\pm z_j\}$ of $\Lambda(u)$ for $A<0$ in the $(s_{\rm L},s_{\rm R})$ plane.
The horizontal axis is $s_{\rm L}=1/\sqrt C$, and the vertical axis is $s_{\rm R}=1/\sqrt{\tilde C}$.}
\label{fig:Lambda-}
\end{figure}

Guided by these numerical observations, we summarize below the ideal ground-state root configurations and the associated phase diagrams in the thermodynamic limit. Finite-size roots may deviate from the ideal positions.
Overall, the bulk zeros of $\Lambda(u)$ form $x$- and $y$-strings, while those of $\bar{\Lambda}(u)$ form $w$-strings. In addition, a small number of zeros lie outside these bulk strings and depend on the boundary parameters.
For $\Lambda(u)$, the phase diagram for $A>0$ is shown in Fig.~\ref{fig:Lambda+}, with the corresponding zero-root configurations in different regions summarized in Table~\ref{tab:Lambda+}.
The results for $A<0$ are presented in Fig.~\ref{fig:Lambda-} and Table~\ref{tab:Lambda-}, respectively.
\begin{longtable}{ccc}
\caption{Ground-state zero-root configurations $\{\pm z_j\}$ of $\Lambda(u)$ in different regions for $A>0$. Here $x_j,y_k\in\mathbb{R}$ parameterize the $x$- and $y$-type bulk strings, respectively. The roots involving $s_{\rm L}$ and $s_{\rm R}$ form boundary strings, while $\pm i z_0$ denote an additional pair of roots with $z_0\in\mathbb{R}$ and $z_0>\tfrac{5}{2}$.}
\label{tab:Lambda+}\\
\toprule
Region & Bulk strings & Remaining roots \\
\midrule
\endfirsthead
\toprule
Region & Bulk strings & Remaining roots \\
\midrule
\endhead
\bottomrule
\endlastfoot
$\mathrm{A}$ &
$\begin{array}{@{}c@{\qquad}l@{}}
x_j,\quad x_j\pm 2i & j=1,\ldots,N \\
y_k\pm\tfrac{5}{2}i & k=1,\ldots,\tfrac{N}{2}-2
\end{array}$ &
$\begin{array}{@{}c@{}}
\pm i\left(s_{\rm L}+\tfrac{3}{2}\right),\quad
\pm i\left(s_{\rm R}+\tfrac{3}{2}\right) \\
\pm iz_0
\end{array}$ \\
\midrule
$\mathrm{B}$ &
$\begin{array}{@{}c@{\qquad}l@{}}
x_j,\quad x_j\pm 2i & j=1,\ldots,N \\
y_k\pm\tfrac{5}{2}i & k=1,\ldots,\tfrac{N}{2}
\end{array}$ &
$\pm i\left[\min\left(s_{\rm L},s_{\rm R}\right)+\tfrac{3}{2}\right]$ \\
\midrule
$\mathrm{C}$ &
$\begin{array}{@{}c@{\qquad}l@{}}
x_j,\quad x_j\pm 2i & j=1,\ldots,N \\
y_k\pm\tfrac{5}{2}i & k=1,\ldots,\tfrac{N}{2}
\end{array}$ &
$\pm iz_0$ \\
\end{longtable}

\begin{longtable}{ccc}
\caption{Ground-state zero-root configurations $\{\pm z_j\}$ of $\Lambda(u)$ in different regions for $A<0$. Here $x_j,y_k\in\mathbb{R}$ parameterize the bulk strings, the roots involving $s_{\rm L}$ and $s_{\rm R}$ form the boundary strings, and $\pm i z_0$ represents an additional root pair with $z_0\in\mathbb{R}$ and $z_0>\tfrac{5}{2}$. The real pair $\pm z_w$, with $z_w\in\mathbb{R}$, is associated with a hole in the $w$-string of $\bar\Lambda(u)$.}
\label{tab:Lambda-}\\
\toprule
Region & Bulk strings & Remaining roots \\
\midrule
\endfirsthead
\toprule
Region & Bulk strings & Remaining roots \\
\midrule
\endhead
\bottomrule
\endlastfoot
$\mathrm{\tilde A}$ &
$\begin{array}{@{}c@{\qquad}l@{}}
x_j,\quad x_j\pm 2i & j=1,\ldots,N \\
y_k\pm\tfrac{5}{2}i & k=1,\ldots,\tfrac{N}{2}-2
\end{array}$ &
$\begin{array}{@{}c@{}}
\pm i\left(s_{\rm L}+\tfrac{3}{2}\right),\quad
\pm i\left(s_{\rm R}+\tfrac{3}{2}\right) \\
\pm z_w
\end{array}$ \\
\midrule
$\mathrm{\tilde B}$ &
$\begin{array}{@{}c@{\qquad}l@{}}
x_j,\quad x_j\pm 2i & j=1,\ldots,N \\
y_k\pm\tfrac{5}{2}i & k=1,\ldots,\tfrac{N}{2}-2
\end{array}$ &
$\begin{array}{@{}c@{}}
\pm i\left[\min\left(s_{\rm L},s_{\rm R}\right)+\tfrac{3}{2}\right] \\
\pm z_w,\quad \pm i z_0
\end{array}$ \\
\midrule
$\mathrm{\tilde C}$ &
$\begin{array}{@{}c@{\qquad}l@{}}
x_j,\quad x_j\pm 2i & j=1,\ldots,N \\
y_k\pm\tfrac{5}{2}i & k=1,\ldots,\tfrac{N}{2}
\end{array}$ &
$\pm z_w$ \\
\end{longtable}

\begin{figure}[htbp]
\centering
\includegraphics[width=0.5\textwidth]{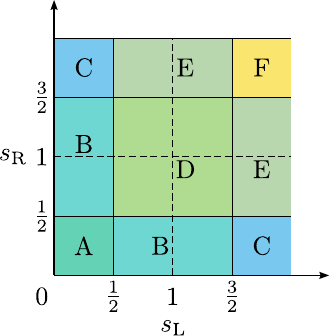}
\caption{Ground-state phase diagram of the zero-root configurations $\{\pm \bar z_j\}$ of $\bar\Lambda(u)$ for $A>0$ in the $(s_{\rm L},s_{\rm R})$ plane.
The horizontal axis is $s_{\rm L}=1/\sqrt C$, and the vertical axis is $s_{\rm R}=1/\sqrt{\tilde C}$.}
\label{fig:Lambdabar+}
\end{figure}

\begin{figure}[htbp]
\centering
\includegraphics[width=0.5\textwidth]{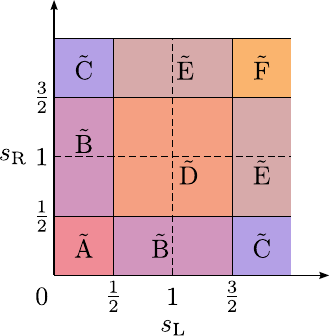}
\caption{Ground-state phase diagram of the zero-root configurations $\{\pm \bar z_j\}$ of $\bar\Lambda(u)$ for $A<0$ in the $(s_{\rm L},s_{\rm R})$ plane.
The horizontal axis is $s_{\rm L}=1/\sqrt C$, and the vertical axis is $s_{\rm R}=1/\sqrt{\tilde C}$.}
\label{fig:Lambdabar-}
\end{figure}

For $\bar\Lambda(u)$, the phase diagram for $A>0$ is shown in Fig.~\ref{fig:Lambdabar+}, with the corresponding zero-root configurations in different regions summarized in Table~\ref{tab:Lambdabar+}.
The results for $A<0$ are presented in Fig.~\ref{fig:Lambdabar-} and Table~\ref{tab:Lambdabar-}, respectively.

\begin{longtable}{ccc}
\caption{Ground-state zero-root configurations $\{\pm\bar z_j\}$ of $\bar\Lambda(u)$ in different regions for $A>0$. Here $w_l\in\mathbb{R}$ parameterizes the $w$-type bulk strings. The boundary variables $s_{\rm L}$ and $s_{\rm R}$ follow the same definitions as those for $\Lambda(u)$, and the roots involving them constitute the boundary strings. The pairs $\pm i\bar z_0$ and $\pm i\bar z_0^{\prime}$ denote additional root pairs, where $\bar z_0,\bar z_0^{\prime}\in\mathbb{R}$ and $\bar z_0,\bar z_0^{\prime}>\tfrac{5}{2}$.}\label{tab:Lambdabar+}\\
\toprule
Region & Bulk strings & Remaining roots \\
\midrule
\endfirsthead
\toprule
Region & Bulk strings & Remaining roots \\
\midrule
\endhead
\midrule
\multicolumn{3}{r}{\textit{Continued on next page}}\\
\endfoot
\bottomrule
\endlastfoot
$\mathrm{A}$ &
$\begin{array}{@{}c@{\qquad}l@{}}
w_l\pm\tfrac{5}{2}i & l=1,\ldots,N-2
\end{array}$ &
$\begin{array}{@{}c@{}}
\pm i\left(s_{\rm L}+1\right),\quad \pm i\left(s_{\rm L}+2\right)\\
\pm i\left(s_{\rm R}+1\right),\quad \pm i\left(s_{\rm R}+2\right)
\end{array}$\\
\midrule
$\mathrm{B}$ &
$\begin{array}{@{}c@{\qquad}l@{}}
w_l\pm\tfrac{5}{2}i & l=1,\ldots,N-2
\end{array}$ &
$\begin{array}{@{}c@{}}
\pm i\left[\min\left(s_{\rm L},s_{\rm R}\right)+1\right],\quad
\pm i\left[\min\left(s_{\rm L},s_{\rm R}\right)+2\right]\\
\pm i\left[\max\left(s_{\rm L},s_{\rm R}\right)+1\right],\quad
\pm i\bar z_0
\end{array}$\\
\midrule
$\mathrm{C}$ &
$\begin{array}{@{}c@{\qquad}l@{}}
w_l\pm\tfrac{5}{2}i & l=1,\ldots,N
\end{array}$ &
$\begin{array}{@{}c@{}}
\pm i\left[\min\left(s_{\rm L},s_{\rm R}\right)+1\right],\quad
\pm i\left[\min\left(s_{\rm L},s_{\rm R}\right)+2\right]
\end{array}$\\
\midrule
$\mathrm{D}$ &
$\begin{array}{@{}c@{\qquad}l@{}}
w_l\pm\tfrac{5}{2}i & l=1,\ldots,N
\end{array}$ &
$\begin{array}{@{}c@{}}
\pm i\left(s_{\rm L}+1\right),\quad
\pm i\left(s_{\rm R}+1\right)
\end{array}$\\
\midrule
$\mathrm{E}$ &
$\begin{array}{@{}c@{\qquad}l@{}}
w_l\pm\tfrac{5}{2}i & l=1,\ldots,N
\end{array}$ &
$\begin{array}{@{}c@{}}
\pm i\left[\min\left(s_{\rm L},s_{\rm R}\right)+1\right],\quad
\pm i\bar z_0
\end{array}$\\
\midrule
$\mathrm{F}$ &
$\begin{array}{@{}c@{\qquad}l@{}}
w_l\pm\tfrac{5}{2}i & l=1,\ldots,N
\end{array}$ &
$\begin{array}{@{}c@{}}
\pm i\bar z_0,\quad
\pm i\bar z_0^{\prime}
\end{array}$\\
\end{longtable}

\begin{longtable}{ccc}
\caption{Ground-state zero-root configurations $\{\pm \bar z_j\}$ of $\bar\Lambda(u)$ in different regions for $A<0$. Here $w_l\in\mathbb{R}$ parameterizes the bulk strings, the roots involving $s_{\rm L}$ and $s_{\rm R}$ correspond to boundary strings, and $\pm i\bar z_0$ denotes an additional root pair with $\bar z_0\in\mathbb{R}$ and $\bar z_0>\tfrac{5}{2}$. The roots $\pm z_w+\tfrac{1}{2}i$ and $\pm z_w-\tfrac{1}{2}i$, with $z_w\in\mathbb{R}$, are associated with a hole in the $w$-string sector.}\label{tab:Lambdabar-}\\
\toprule
Region & Bulk strings & Remaining roots \\
\midrule
\endfirsthead
\toprule
Region & Bulk strings & Remaining roots \\
\midrule
\endhead
\midrule
\multicolumn{3}{r}{\textit{Continued on next page}}\\
\endfoot
\bottomrule
\endlastfoot
$\mathrm{\tilde A}$ &
$\begin{array}{@{}c@{\qquad}l@{}}
w_l\pm\tfrac{5}{2}i & l=1,\ldots,N-4
\end{array}$ &
$\begin{array}{@{}c@{}}
\pm i\left(s_{\rm L}+1\right),\quad \pm i\left(s_{\rm L}+2\right)\\
\pm i\left(s_{\rm R}+1\right),\quad \pm i\left(s_{\rm R}+2\right)\\
\pm z_w +\tfrac 12i,\quad \pm z_w -\tfrac 12i
\end{array}$\\
\midrule
$\mathrm{\tilde B}$ &
$\begin{array}{@{}c@{\qquad}l@{}}
w_l\pm\tfrac{5}{2}i & l=1,\ldots,N-4
\end{array}$ &
$\begin{array}{@{}c@{}}
\pm i\left[\min\left(s_{\rm L},s_{\rm R}\right)+1\right],\quad
\pm i\left[\min\left(s_{\rm L},s_{\rm R}\right)+2\right]\\
\pm i\left[\max\left(s_{\rm L},s_{\rm R}\right)+1\right],\quad
\pm i\bar z_0\\
\pm z_w +\tfrac 12i,\quad \pm z_w -\tfrac 12i
\end{array}$\\
\midrule
$\mathrm{\tilde C}$ &
$\begin{array}{@{}c@{\qquad}l@{}}
w_l\pm\tfrac{5}{2}i & l=1,\ldots,N-2
\end{array}$ &
$\begin{array}{@{}c@{}}
\pm i\left[\min\left(s_{\rm L},s_{\rm R}\right)+1\right],\quad
\pm i\left[\min\left(s_{\rm L},s_{\rm R}\right)+2\right]\\
\pm z_w +\tfrac 12i,\quad \pm z_w -\tfrac 12i
\end{array}$\\
\midrule
$\mathrm{\tilde D}$ &
$\begin{array}{@{}c@{\qquad}l@{}}
w_l\pm\tfrac{5}{2}i & l=1,\ldots,N-2
\end{array}$ &
$\begin{array}{@{}c@{}}
\pm i\left(s_{\rm L}+1\right),\quad
\pm i\left(s_{\rm R}+1\right)\\
\pm z_w +\tfrac 12i,\quad \pm z_w -\tfrac 12i
\end{array}$\\
\midrule
$\mathrm{\tilde E}$ &
$\begin{array}{@{}c@{\qquad}l@{}}
w_l\pm\tfrac{5}{2}i & l=1,\ldots,N-2
\end{array}$ &
$\begin{array}{@{}c@{}}
\pm i\left[\min\left(s_{\rm L},s_{\rm R}\right)+1\right],\quad
\pm i\bar z_0\\
\pm z_w +\tfrac 12i,\quad \pm z_w -\tfrac 12i
\end{array}$\\
\midrule
$\mathrm{\tilde F}$ &
$\begin{array}{@{}c@{\qquad}l@{}}
w_l\pm\tfrac{5}{2}i & l=1,\ldots,N
\end{array}$ &
$\begin{array}{@{}c@{}}
\pm z_w +\tfrac 12i,\quad \pm z_w -\tfrac 12i
\end{array}$\\
\end{longtable}

Fig.~\ref{fig:ground-representatives} presents four representative ground-state zero-root configurations from the classifications in Tables~\ref{tab:Lambda+}--\ref{tab:Lambdabar-}. 
Panels (a), (b), (c), and (d) correspond to the region pairs $(\mathrm A,\mathrm B)$, $(\mathrm C,\mathrm D)$, $(\mathrm C,\mathrm F)$, and $(\tilde{\mathrm B},\tilde{\mathrm E})$ for $(\Lambda,\bar\Lambda)$, respectively.
Panels (a), (b), and (c) belong to the $A>0$ classifications, whereas panel (d) belongs to the $A<0$ classifications.
A quantitative comparison of the numerical boundary roots with their ideal positions is given in Appendix~\ref{app:numerical-boundary-roots}.

In panel (d), where $\bar\Lambda$ is in region $\tilde{\mathrm E}$, finite-size effects cause the roots to deviate appreciably from the ideal quartet $\pm z_w\pm\tfrac{i}{2}$ listed in Table~\ref{tab:Lambdabar-}.
This structure is more clearly resolved in the excited-state configuration shown in Fig.~\ref{fig:W-excitation}, and the associated $W$-type excitation is analyzed in Sec.~\ref{sec:W-type Excitation}.
The regularity argument fixing the common real center and imaginary offsets of this ideal configuration is given in Appendix~\ref{app:W-center-locking}.

\begin{figure}[htbp]
\centering
\includegraphics[width=0.7\textwidth]{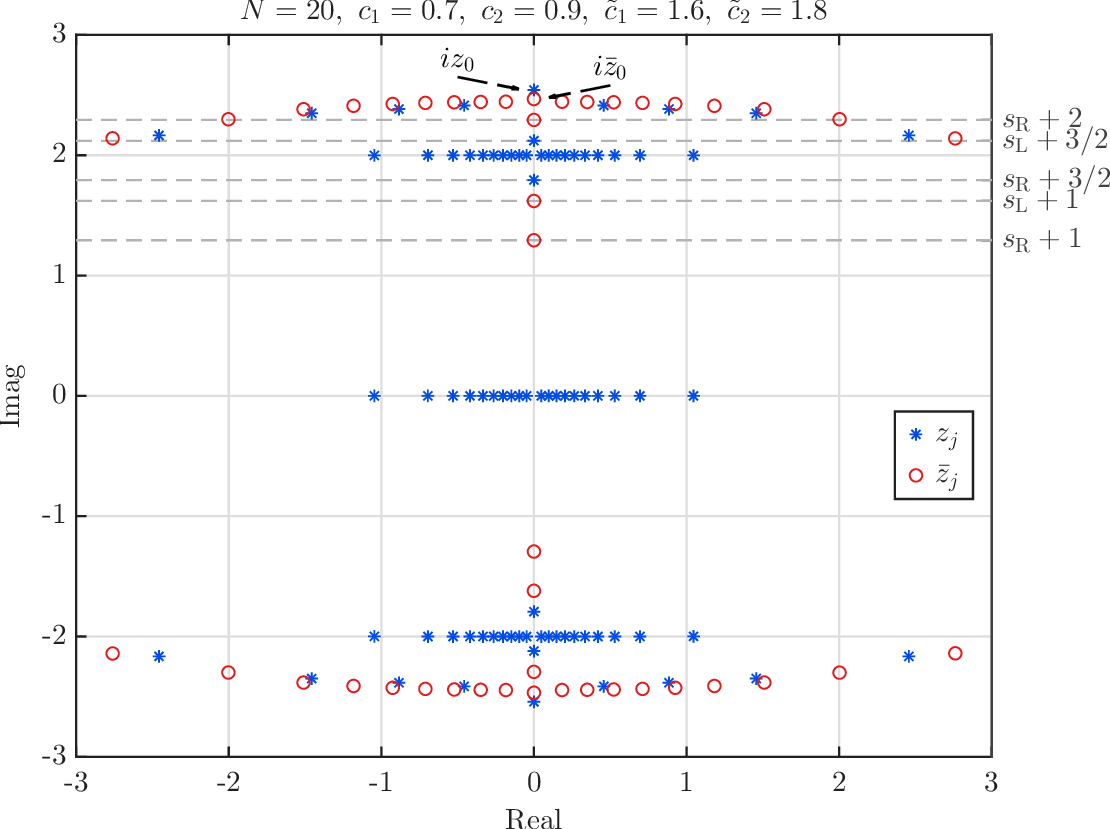}
\par\smallskip
(a) $\Lambda:\mathrm A,\quad \bar\Lambda:\mathrm B$
\par\vspace{0.5cm}
\includegraphics[width=0.7\textwidth]{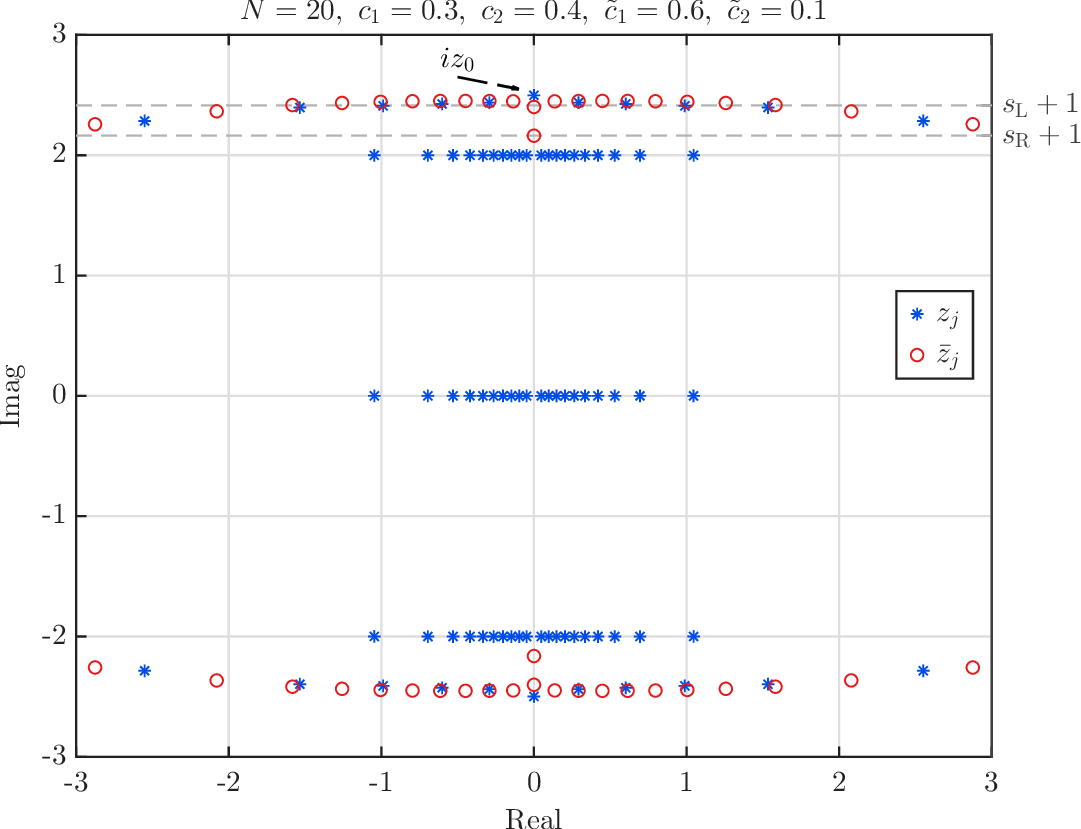}
\par\smallskip
(b) $\Lambda:\mathrm C,\quad \bar\Lambda:\mathrm D$
\par
\caption{Ground-state zero-root distributions in representative regions classified in Tables~\ref{tab:Lambda+}--\ref{tab:Lambdabar-}.
The $x$-strings of $\Lambda(u)$ are clearly visible as blue stars along the real axis and with imaginary parts close to $\pm2$.
The blue stars and red circles with imaginary parts close to $\pm\tfrac52$ form the $y$-strings of $\Lambda(u)$ and the $w$-strings of $\bar\Lambda(u)$, respectively.
Horizontal dashed lines indicate the ideal imaginary parts of the boundary-string roots in the upper half-plane.
Arrows indicate additional roots and, where present, hole-related roots.
A quantitative comparison of the ideal and numerical boundary-root positions, rounded to five decimal places, is given in Appendix~\ref{app:numerical-boundary-roots}.}
\label{fig:ground-representatives}
\end{figure}
\begin{figure}[htbp]
\ContinuedFloat
\centering
\includegraphics[width=0.63\textwidth]{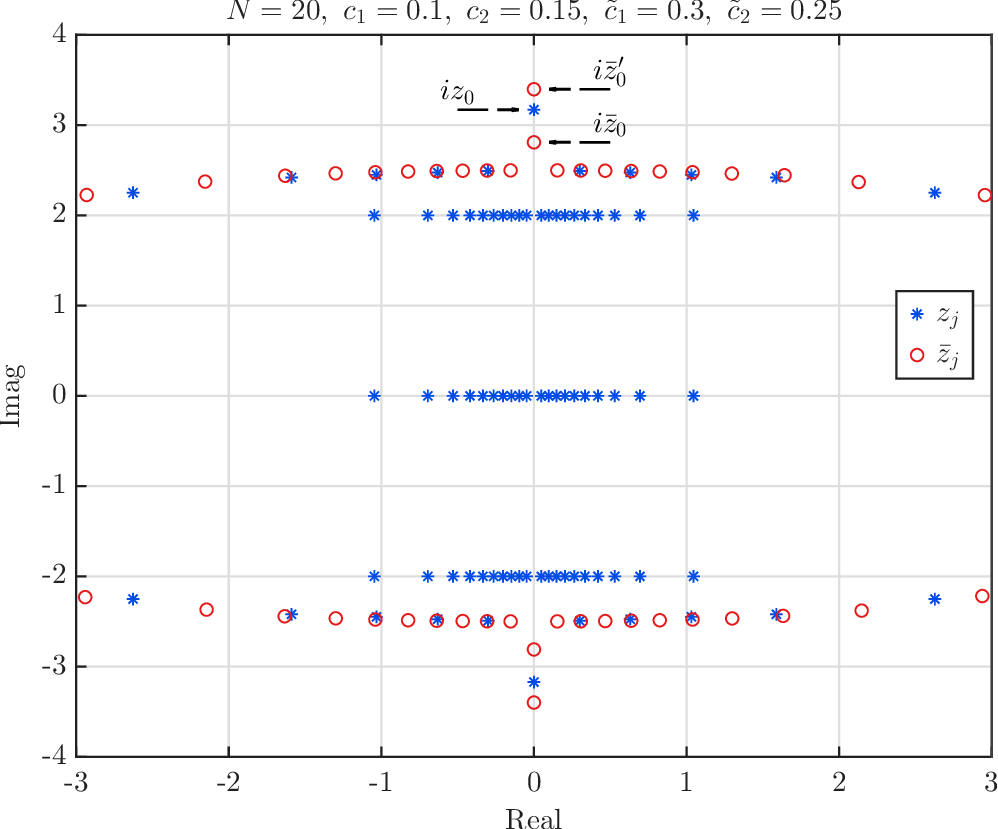}
\par\smallskip
(c) $\Lambda:\mathrm C,\quad \bar\Lambda:\mathrm F$
\par\vspace{0.5cm}\qquad
\includegraphics[width=0.7\textwidth]{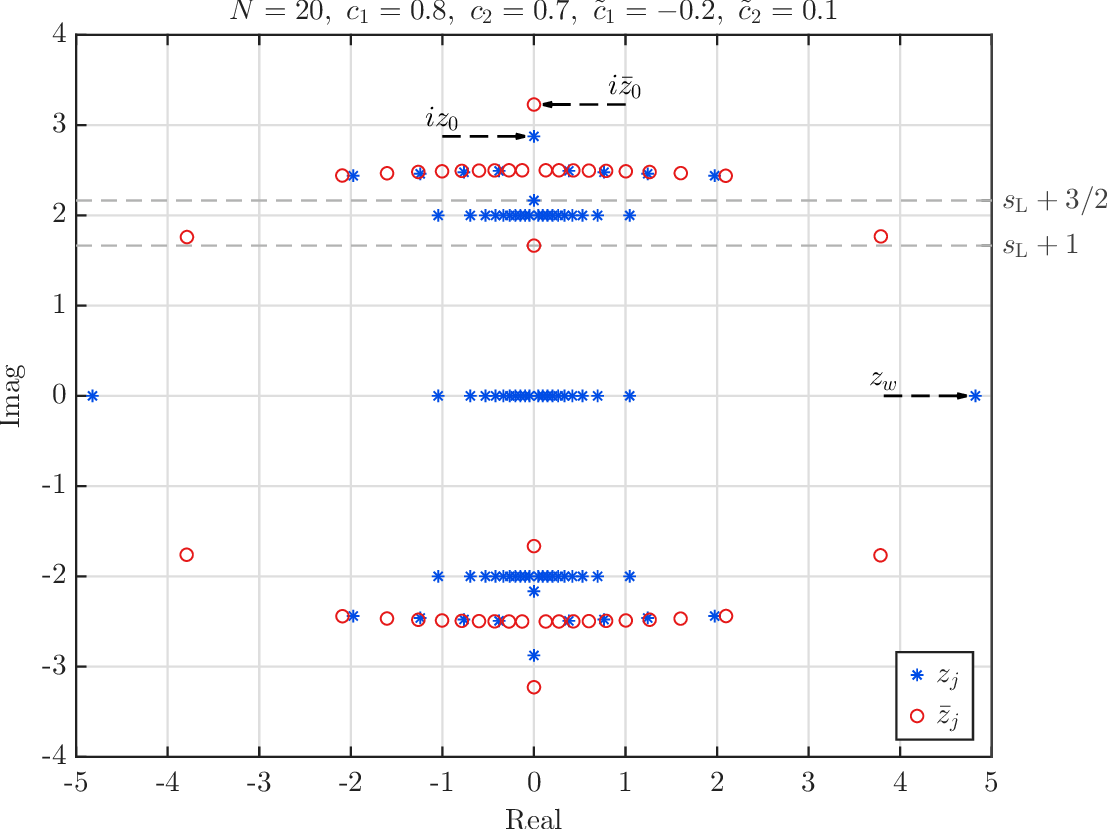}
\par\smallskip
(d) $\Lambda:\tilde{\mathrm B},\quad \bar\Lambda:\tilde{\mathrm E}$
\par
\caption[]{Continued.}
\end{figure}

\subsection{Root Densities and Ground-State Energy}
From Tables~\ref{tab:Lambda+}--\ref{tab:Lambdabar-}, two distinct features can be identified.
First, the sign of $A$ affects the presence or absence of a hole in the $w$-string sector, while preserving the bulk-string forms.
Second, the finite purely imaginary roots are associated with the boundary variables $s_{\rm L}$ and $s_{\rm R}$.
For each boundary, the numerical ground-state configurations contain at most one pair of such roots in $\Lambda(u)$ and at most two pairs in $\bar\Lambda(u)$.
As illustrated by the phase diagrams in Figs.~\ref{fig:Lambda+} and \ref{fig:Lambdabar+}, the two boundaries play symmetric roles in the organization of these root configurations.

To simplify the analysis of the boundary roots, we first focus on the $A>0$ ground state, whose $w$-string distribution contains no holes.
Taking advantage of the symmetry between the two boundaries, we further choose $s_{\rm R}>\tfrac{3}{2}$, so that only the left boundary can contribute boundary strings.

This restriction selects regions $\mathrm B$ and $\mathrm C$ in Fig.~\ref{fig:Lambda+}, and regions $\mathrm C$, $\mathrm E$, and $\mathrm F$ in Fig.~\ref{fig:Lambdabar+}.

Although the boundary-string content changes between these regions, Tables~\ref{tab:Lambda+} and \ref{tab:Lambdabar+} show that the numbers of finite purely imaginary roots outside the bulk strings remain unchanged: one pair in $\Lambda(u)$ and two pairs in $\bar\Lambda(u)$.
These pairs may represent boundary strings or additional roots, depending on the region.
We therefore parameterize them uniformly as $\pm i\alpha$ in $\Lambda(u)$ and $\pm i\beta_1$, $\pm i\beta_2$ in $\bar\Lambda(u)$.
Their positions in the selected ground-state configurations satisfy

\begin{equation}
\alpha>\tfrac{3}{2},\qquad \beta_1>1,\qquad \beta_2>1.
\label{eq:boundary-root-range}
\end{equation}
The other zeros form the corresponding bulk strings.

Starting from this representative ground-state configuration, we derive the zero-root distributions in the thermodynamic limit and determine analytically when the finite purely imaginary roots are fixed by the boundary parameters.

We use the analytical method developed in Refs.~\cite{qiao2,qiao3,qiao1} to calculate the ground-state energy.
The first step is to derive continuum equations from the operator identities in the thermodynamic limit.
At finite system size, the identities in Eqs.~\eqref{eq:I}--\eqref{eq:III} hold only at the discrete inhomogeneity points.
In the thermodynamic limit $N\rightarrow\infty$, these inhomogeneous parameters can be taken to form a continuous distribution.

\begin{figure}[tbp]
\centering
\begin{minipage}{0.48\textwidth}
\centering
\includegraphics[width=\linewidth]{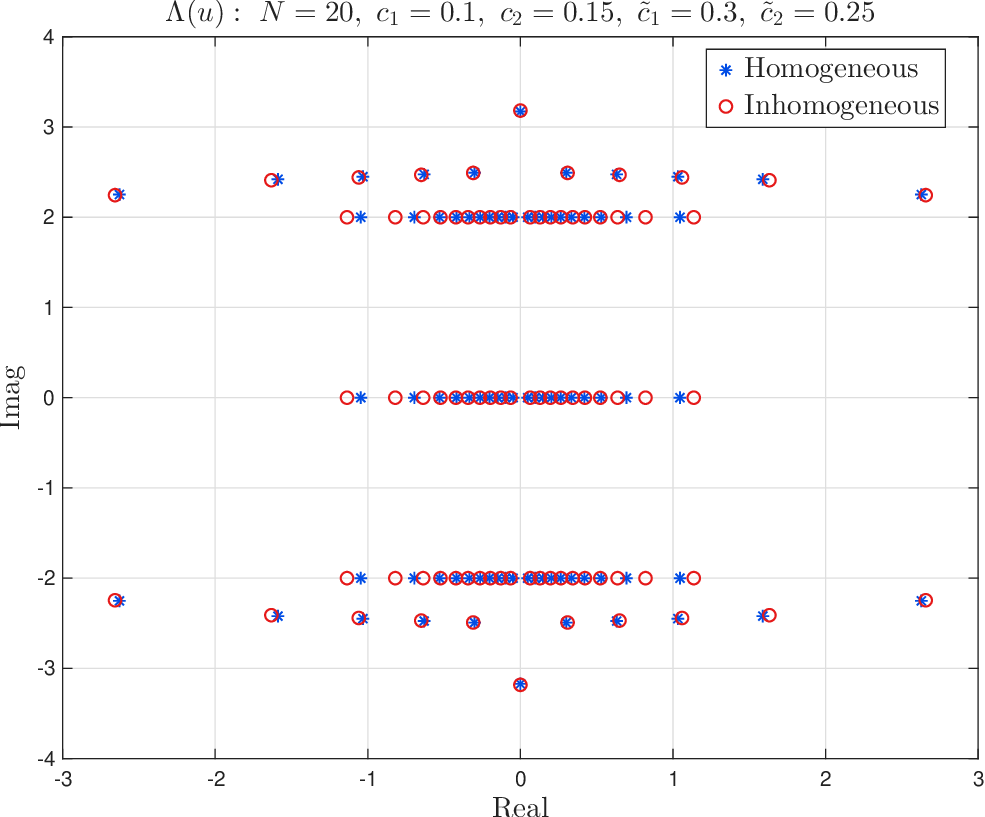}
\par\smallskip
(a) $\Lambda(u)$
\end{minipage}
\hfill
\begin{minipage}{0.48\textwidth}
\centering
\includegraphics[width=\linewidth]{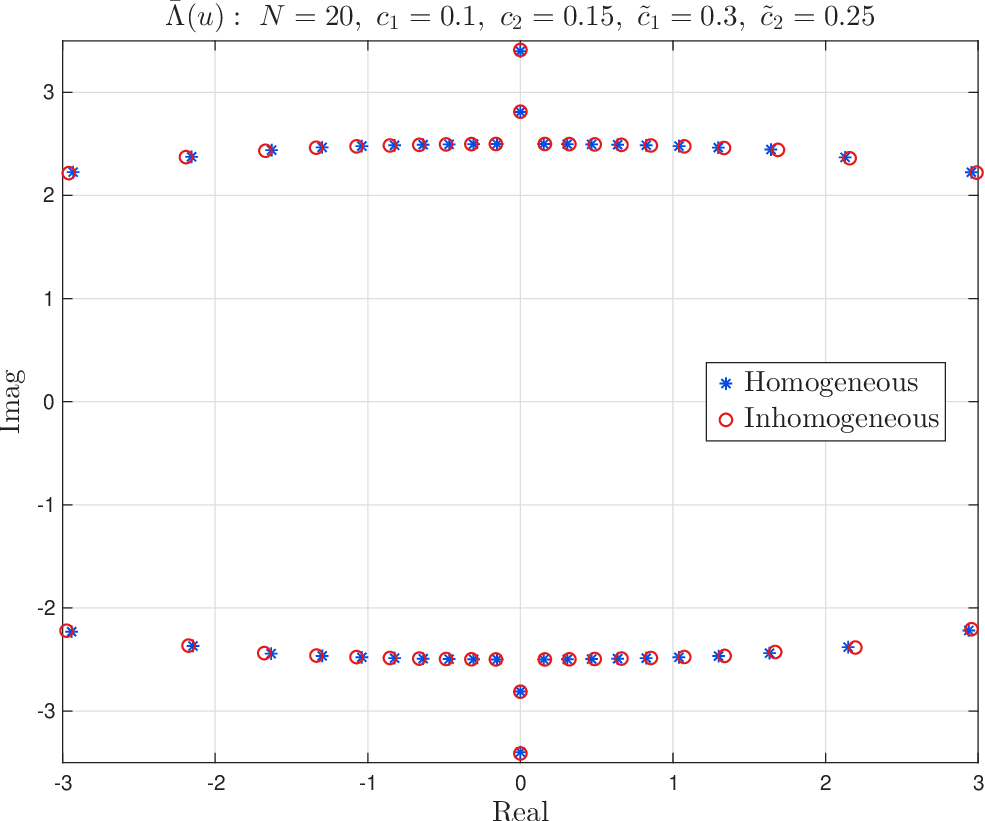}
\par\smallskip
(b) $\bar\Lambda(u)$
\end{minipage}

\medskip
\begin{minipage}{0.48\textwidth}
\centering
\includegraphics[width=\linewidth]{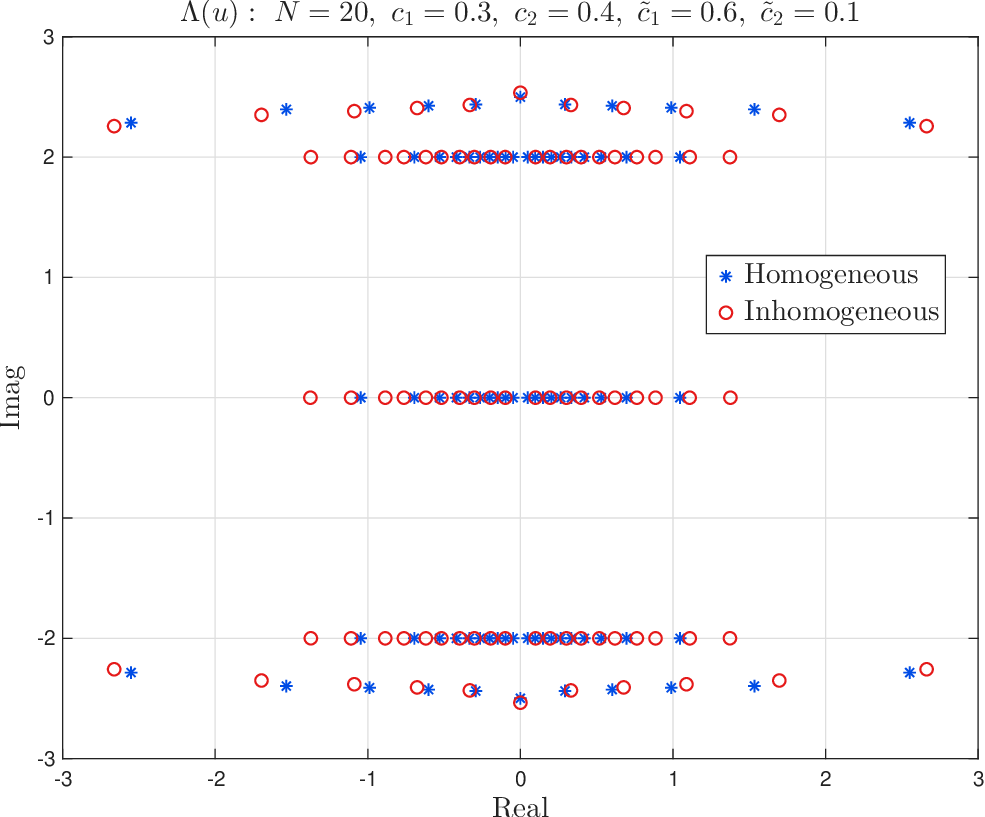}
\par\smallskip
(c) $\Lambda(u)$
\end{minipage}
\hfill
\begin{minipage}{0.48\textwidth}
\centering
\includegraphics[width=\linewidth]{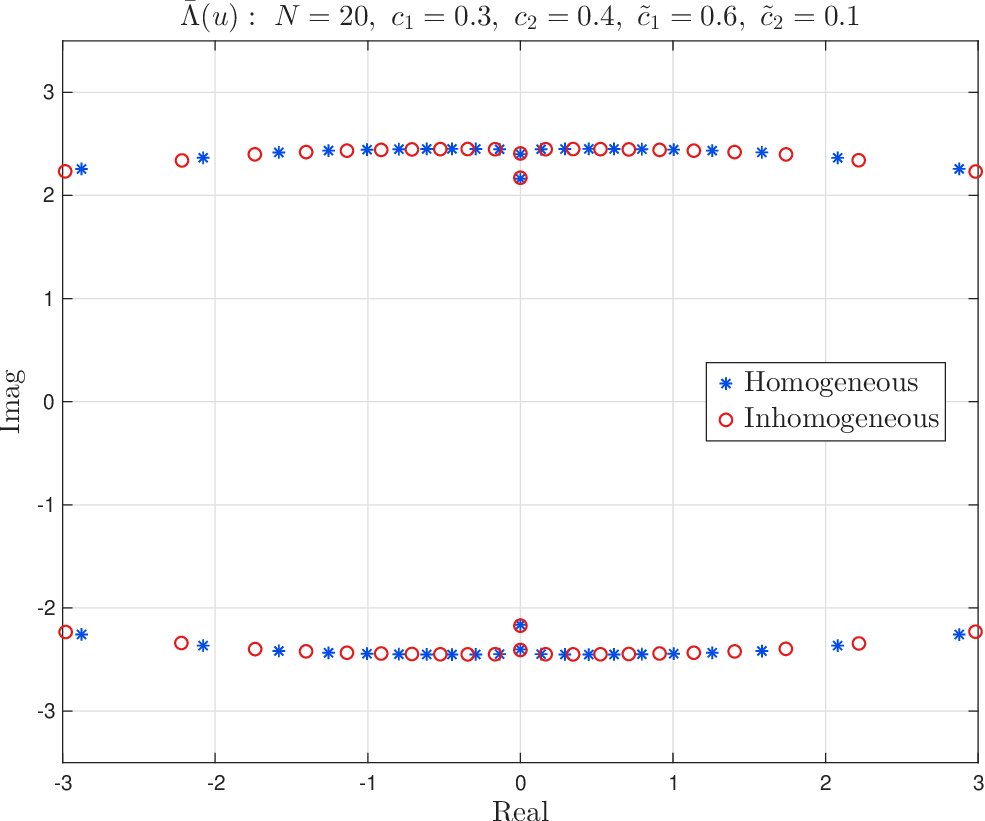}
\par\smallskip
(d) $\bar\Lambda(u)$
\end{minipage}
\caption{Comparison of the homogeneous and inhomogeneous zero-root distributions of $\Lambda(u)$ and $\bar{\Lambda}(u)$. Blue stars represent the homogeneous roots, whereas red circles represent the inhomogeneous roots. The inhomogeneous parameters are $\hat{\theta}_j=0.05\left(j-\tfrac{N+1}{2}\right)$ in panels (a) and (b), and $\hat{\theta}_j=0.1\left(j-\tfrac{N+1}{2}\right)$ in panels (c) and (d), with $j=1,\ldots,N$.}
\label{fig:imaginary-inhomogeneities}
\end{figure}

Before carrying out the continuum analysis, we first check whether the ground-state string structure is preserved after introducing nonzero inhomogeneities.
To this end, we take the inhomogeneities to be purely imaginary while keeping the boundary parameters fixed.
As shown in Fig.~\ref{fig:imaginary-inhomogeneities}, the homogeneous and inhomogeneous roots of both transfer matrices exhibit the same string structure for the examples considered.
This confirms that the same string configuration can be used in the continuum treatment.
We therefore can write
\begin{equation}
\theta_j=i\hat{\theta}_j,\qquad \hat{\theta}_j\in\mathbb{R},
\qquad j=1,\ldots,N,
\end{equation}
and allow the real variables $\{\hat{\theta}_j\}$ to become continuously distributed in the thermodynamic limit.

The second step is to approximate the continuous logarithmic derivatives by finite differences between neighboring inhomogeneous points.
Before applying the finite-difference approximation, we first combine Eqs.~\eqref{eq:II} and \eqref{eq:III} through their product and ratio, so that the odd contributions present in the individual equations cancel.
After substituting the zero-root parametrizations and taking the logarithmic derivatives, two elementary functions naturally appear,
\begin{equation}
a_n(v)=\frac{2n}{v^2+n^2},
\qquad
b_n(v)=\frac{2v}{v^2+n^2}.
\label{eq:ab-kernels}
\end{equation}
In the thermodynamic limit, the local root densities are related to the spacings between neighboring roots.
The density of the inhomogeneous parameters is defined analogously.
Introducing $\mu_j$ with $\mu\in\{x,y,w,\hat{\theta}\}$, we have
\begin{equation}
N\rho_\mu\!\left(\mu_j\right)
\left(
\mu_{j+1}-\mu_j
\right)
\sim 1,
\qquad
\mu\in\{x,y,w,\hat{\theta}\},
\label{eq:root-count-density}
\end{equation}
and hence
\begin{equation}
\rho_\mu\!\left(\mu_j\right)
\sim
\frac{1}{
N\left(
\mu_{j+1}-\mu_j
\right)},
\qquad
\mu\in\{x,y,w,\hat{\theta}\}.
\label{eq:density-definition}
\end{equation}
Applying this finite-difference procedure to the logarithms of Eq.~\eqref{eq:I} and the product and ratio of Eqs.~\eqref{eq:II} and \eqref{eq:III}, and neglecting the resulting $O(\tfrac{1}{N})$ approximation error, we obtain the coupled density equations
\begin{equation}
\begin{aligned}
&N\int_{-\infty}^{\infty}\rho_x(u)
\Big[a_{\frac12}(v-u)+a_{\frac32}(v-u)+a_{\frac72}(v-u)\Big]\,du
\\
&\quad
+N\int_{-\infty}^{\infty}\rho_y(u)
\Big[a_1(v-u)+a_4(v-u)\Big]\,du
\\
&=
2N\int_{-\infty}^{\infty}\rho_{\hat{\theta}}(u)
\Big[a_1(v-u)+a_3(v-u)\Big]\,du
\\
&\quad
+a_1(v)+a_3(v)-a_{\frac12}(v)-a_{\frac32}(v)
\\
&\quad
+a_{s_{\rm L}}(v)+a_{s_{\rm R}}(v)
-a_{\alpha-\frac32}(v)-a_{\alpha+\frac32}(v).
\end{aligned}
\label{eq:density-I}
\end{equation}

\begin{equation}
\begin{aligned}
&N\int_{-\infty}^{\infty}\rho_x(u)
\Big[
2a_{\frac12}(v-u)+2a_{\frac32}(v-u)
+a_{\frac52}(v-u)+a_{\frac72}(v-u)
\Big]\,du
\\
&\quad
+N\int_{-\infty}^{\infty}\rho_y(u)
\Big[
a_1(v-u)+a_2(v-u)+a_3(v-u)+a_4(v-u)
\Big]\,du
\\
&\quad
-N\int_{-\infty}^{\infty}\rho_w(u)
\Big[a_{\frac32}(v-u)+a_{\frac72}(v-u)\Big]\,du
\\
&=
2N\int_{-\infty}^{\infty}\rho_{\hat{\theta}}(u)
\Big[2a_1(v-u)+a_3(v-u)\Big]\,du
\\
&\quad
+3a_1(v)+a_3(v)-2a_{\frac12}(v)
-a_{\frac32}(v)-a_{\frac52}(v)
\\
&\quad
-a_{\alpha-\frac32}(v)-a_{\alpha-\frac12}(v)
-a_{\alpha+\frac12}(v)-a_{\alpha+\frac32}(v)
\\
&\quad
+a_{\beta_1-1}(v)+a_{\beta_1+1}(v)
+a_{\beta_2-1}(v)+a_{\beta_2+1}(v).
\end{aligned}
\label{eq:density-II}
\end{equation}

\begin{equation}
\begin{aligned}
&N\int_{-\infty}^{\infty}\rho_x(u)
\Big[a_{\frac52}(v-u)+a_{\frac72}(v-u)\Big]\,du
\\
&\quad
-N\int_{-\infty}^{\infty}\rho_y(u)
\Big[
a_1(v-u)+a_2(v-u)-a_3(v-u)-a_4(v-u)
\Big]\,du
\\
&\quad
-N\int_{-\infty}^{\infty}\rho_w(u)
\Big[a_{\frac72}(v-u)-a_{\frac32}(v-u)\Big]\,du
\\
&=
2N\int_{-\infty}^{\infty}
\rho_{\hat{\theta}}(u)a_3(v-u)\,du
\\
&\quad
+a_1(v)+a_3(v)-a_{\frac32}(v)-a_{\frac52}(v)
\\
&\quad
+a_{\alpha-\frac32}(v)+a_{\alpha-\frac12}(v)
-a_{\alpha+\frac12}(v)-a_{\alpha+\frac32}(v)
\\
&\quad
-a_{\beta_1-1}(v)-a_{\beta_2-1}(v)
+a_{\beta_1+1}(v)+a_{\beta_2+1}(v).
\end{aligned}
\label{eq:density-III}
\end{equation}

The third step is to perform the Fourier transformation of the coupled density equations.
We use the convention
\begin{equation}
\tilde f(\omega)
=
\int_{-\infty}^{\infty}
f(v)e^{i\omega v}\,dv,
\qquad
f(v)
=
\frac{1}{2\pi}
\int_{-\infty}^{\infty}
\tilde f(\omega)e^{-i\omega v}\,d\omega.
\label{eq:Fourier-transform}
\end{equation}
For $n>0$, the Fourier transforms of the functions defined above are
\begin{equation}
\tilde a_n(\omega)
=
2\pi e^{-n|\omega|},
\qquad
\tilde b_n(\omega)
=
2\pi i\,\operatorname{sgn}(\omega)e^{-n|\omega|}.
\label{eq:Fourier-ab}
\end{equation}
Introducing
\begin{equation}
q=e^{-|\omega|/2},
\label{eq:q-definition}
\end{equation}
the three coupled integral equations become
\begin{align}
&N\left(q+q^3+q^7\right)\tilde{\rho}_x
+N\left(q^2+q^8\right)\tilde{\rho}_y
=
2N\left(q^2+q^6\right)\tilde{\rho}_{\hat{\theta}}
+q^2+q^6-q-q^3
\nonumber\\
&\hspace{34mm}
+q^{2s_{\rm L}}+q^{2s_{\rm R}}
-q^{2\alpha-3}-q^{2\alpha+3},
\label{eq:Fourier-density-I}
\\[1ex]
&N\left(2q+2q^3+q^5+q^7\right)\tilde{\rho}_x
+N\left(q^2+q^4+q^6+q^8\right)\tilde{\rho}_y
-N\left(q^3+q^7\right)\tilde{\rho}_w
\nonumber\\
&\qquad=
2N\left(2q^2+q^6\right)\tilde{\rho}_{\hat{\theta}}
+3q^2+q^6-2q-q^3-q^5
\nonumber\\
&\hspace{22mm}
-q^{2\alpha-3}-q^{2\alpha-1}
-q^{2\alpha+1}-q^{2\alpha+3}
\nonumber\\
&\hspace{22mm}
+q^{2\beta_1-2}+q^{2\beta_1+2}
+q^{2\beta_2-2}+q^{2\beta_2+2},
\label{eq:Fourier-density-II}
\\[1ex]
&-N\left(q^5+q^7\right)\tilde{\rho}_x
+N\left(q^2+q^4-q^6-q^8\right)\tilde{\rho}_y
+N\left(q^7-q^3\right)\tilde{\rho}_w
\nonumber\\
&\qquad=
-2Nq^6\tilde{\rho}_{\hat{\theta}}
-q^2-q^6+q^3+q^5
\nonumber\\
&\hspace{22mm}
-q^{2\alpha-3}-q^{2\alpha-1}
+q^{2\alpha+1}+q^{2\alpha+3}
\nonumber\\
&\hspace{22mm}
+q^{2\beta_1-2}-q^{2\beta_1+2}
+q^{2\beta_2-2}-q^{2\beta_2+2}.
\label{eq:Fourier-density-III}
\end{align}
Solving Eqs.~\eqref{eq:Fourier-density-I}--\eqref{eq:Fourier-density-III} while retaining the inhomogeneity distribution $\tilde{\rho}_{\hat{\theta}}(\omega)$, we obtain
\begin{equation}
\tilde{\rho}_x(\omega)
=
\frac{2q}{1+q^2}\tilde{\rho}_{\hat{\theta}}(\omega)
-\frac{(q-1)^2}{N(1+q^2)}.
\label{eq:rho-x-inhomogeneous}
\end{equation}
\begin{equation}
\begin{aligned}
\tilde{\rho}_y(\omega)
={}&
\frac{2q^4}{(1+q^2)(1+q^6)}\tilde{\rho}_{\hat{\theta}}(\omega)
+\frac{
q^7-q^6+q^5+q^4-q^2-1
}{
N(1+q^2)(1+q^6)
}
\\
&+
\frac{
q^{2s_{\rm L}-2}+q^{2s_{\rm R}-2}
-(1+q^6)q^{2\alpha-5}
}{
N(1+q^6)
}.
\end{aligned}
\label{eq:rho-y-inhomogeneous}
\end{equation}
\begin{equation}
\begin{aligned}
\tilde{\rho}_w(\omega)
={}&
\frac{2q^3}{1+q^6}\tilde{\rho}_{\hat{\theta}}(\omega)
+\frac{
q^4+q^3-q-1
}{
N(1+q^6)
}
\\
&+
\frac{
(1+q^2)
\left(
q^{2s_{\rm L}-3}+q^{2s_{\rm R}-3}
\right)
-(1+q^6)
\left(
q^{2\beta_1-5}+q^{2\beta_2-5}
\right)
}{
N(1+q^6)
}.
\end{aligned}
\label{eq:rho-w-inhomogeneous}
\end{equation}
For the root configuration considered here, the inhomogeneity distribution enters explicitly through the leading bulk terms, while the finite-root and boundary contributions retain their dependence on $\alpha$, $\beta_{1,2}$ and $s_{\rm L,R}$.
In the homogeneous limit, the density of the inhomogeneous parameters becomes
\begin{equation}
\rho_{\hat{\theta}}(v)=\delta(v),
\qquad
\tilde{\rho}_{\hat{\theta}}(\omega)=1.
\label{eq:homogeneous-limit}
\end{equation}
Substituting Eq.~\eqref{eq:homogeneous-limit} into Eqs.~\eqref{eq:rho-x-inhomogeneous}--\eqref{eq:rho-w-inhomogeneous} gives
\begin{equation}
\tilde{\rho}_x(\omega)
=
\frac{2q}{1+q^2}
-\frac{(q-1)^2}{N(1+q^2)}.
\label{eq:rho-x}
\end{equation}
\begin{equation}
\begin{aligned}
\tilde{\rho}_y(\omega)
={}&
\frac{2q^4}{(1+q^2)(1+q^6)}
+\frac{
q^7-q^6+q^5+q^4-q^2-1
}{
N(1+q^2)(1+q^6)
}
\\
&+
\frac{
q^{2s_{\rm L}-2}+q^{2s_{\rm R}-2}
-(1+q^6)q^{2\alpha-5}
}{
N(1+q^6)
}.
\end{aligned}
\label{eq:rho-y}
\end{equation}
\begin{equation}
\begin{aligned}
\tilde{\rho}_w(\omega)
={}&
\frac{2q^3}{1+q^6}
+\frac{
q^4+q^3-q-1
}{
N(1+q^6)
}
\\
&+
\frac{
(1+q^2)
\left(
q^{2s_{\rm L}-3}+q^{2s_{\rm R}-3}
\right)
-(1+q^6)
\left(
q^{2\beta_1-5}+q^{2\beta_2-5}
\right)
}{
N(1+q^6)
}.
\end{aligned}
\label{eq:rho-w}
\end{equation}
The $O(1)$ terms give the bulk root distributions:
\begin{equation}
\begin{aligned}
\tilde{\rho}_x^{(0)}(\omega)
&=
\operatorname{sech}\left(\frac{\omega}{2}\right),
&\qquad
\rho_x^{(0)}(v)
&=
\operatorname{sech}(\pi v),
\\
\tilde{\rho}_w^{(0)}(\omega)
&=
\operatorname{sech}\left(\frac{3\omega}{2}\right),
&
\rho_w^{(0)}(v)
&=
\frac{1}{3}
\operatorname{sech}\left(\frac{\pi v}{3}\right),
\\
\tilde{\rho}_y^{(0)}(\omega)
&=
\frac{1}{2}
\operatorname{sech}\left(\frac{\omega}{2}\right)
\operatorname{sech}\left(\frac{3\omega}{2}\right),
&
\rho_y^{(0)}(v)
&=
\frac{
\tfrac{2}{\sqrt{3}}\sinh\left(\tfrac{2\pi v}{3}\right)-v
}{
3\sinh(\pi v)
}.
\end{aligned}
\label{eq:bulk-densities}
\end{equation}
In particular,
\begin{equation}
\tilde{\rho}_y^{(0)}(\omega)
=
\frac{1}{2}
\tilde{\rho}_x^{(0)}(\omega)
\tilde{\rho}_w^{(0)}(\omega).
\label{eq:bulk-density-relation}
\end{equation}

The bulk densities $\rho_x^{(0)}$, $\rho_y^{(0)}$, and $\rho_w^{(0)}$ of the present open chain with non-diagonal boundaries correspond to the periodic-chain Bethe-string center densities $\sigma^{(1)}$, $\sigma^{(3)}$, and $\sigma^{(2)}$, respectively, differing by a factor of two due to normalization~\cite{martins}.
The zero-root strings and their Bethe-root counterparts also differ in their imaginary parts: the $x$-strings lie at $0,\pm2$, whereas the corresponding Bethe roots are real; the $y$-strings lie at $\pm\tfrac52$, compared with $\pm\tfrac12$ for the corresponding Bethe roots; the $w$-strings lie at $\pm\tfrac52$, whereas their Bethe-root counterparts are real.

According to Eq.~\eqref{eq:Hamiltonian-general}, the energy can be expressed as
\begin{equation}
E=
\left.
\frac{3}{2}\frac{d}{du}\ln\Lambda(u)
\right|_{u=0,\{\theta_j\}=0}
=
\frac{9}{2}\sum_{j=1}^{2N+1}
\frac{1}{z_j^2+\tfrac94}.
\label{eq:energy}
\end{equation}

In the thermodynamic limit, the extensive parts of the zero roots are described by the $x$- and $y$-string densities.
Denoting the finite zero roots of $\Lambda(u)$ that do not belong to these bulk strings by
\begin{equation}
\{z_r^{\rm f}\}_{r=1}^{M_{\rm f}},
\label{eq:finite-zero-roots}
\end{equation}
Eq.~\eqref{eq:energy} can be written as
\begin{equation}
\begin{aligned}
E
={}&
\frac{3N}{4}\int_{-\infty}^{\infty}\rho_x(v)
\left[
a_{\frac32}(v)+a_{\frac72}(v)-a_{\frac12}(v)
\right]\,dv
\\
&+
\frac{3N}{4}\int_{-\infty}^{\infty}\rho_y(v)
\left[
a_4(v)-a_1(v)
\right]\,dv
+
\frac{9}{2}\sum_{r=1}^{M_{\rm f}}
\frac{1}{(z_r^{\rm f})^2+\tfrac94}
\\
={}&
\frac{3N}{2}\int_{0}^{\infty}
\left[
(q^3+q^7-q)\tilde{\rho}_x(\omega)
+(q^8-q^2)\tilde{\rho}_y(\omega)
\right]\,d\omega
+
\frac{9}{2}\sum_{r=1}^{M_{\rm f}}
\frac{1}{(z_r^{\rm f})^2+\tfrac94}.
\end{aligned}
\label{eq:energy-thermodynamic}
\end{equation}
For the ground-state configuration considered above, the only finite zero root of $\Lambda(u)$ outside the bulk strings is
\begin{equation}
z_1^{\rm f}=i\alpha.
\end{equation}
Substituting this finite root together with the ground-state densities in Eqs.~\eqref{eq:rho-x} and \eqref{eq:rho-y} into Eq.~\eqref{eq:energy-thermodynamic}, we obtain
\begin{equation}
\begin{aligned}
E_g
={}&
N\left(
3-4\ln2-\frac{2\sqrt3\,\pi}{9}
\right)
-3-2\ln2
+\frac{(9+\sqrt3)\pi}{9}
+\sqrt3\ln(2+\sqrt3)
\\
&-\frac{3}{2}\sqrt C-\frac{3}{2}\sqrt{\tilde C}
+\frac{3}{2}\int_{-\infty}^{\infty}
\frac{
e^{-|\omega|/\sqrt C}
+e^{-|\omega|/\sqrt{\tilde C}}
}{
1+e^{3|\omega|}
}\,d\omega.
\end{aligned}
\label{eq:ground-state-energy}
\end{equation}
The $\alpha$-dependent terms cancel exactly, leaving no explicit $\alpha$ dependence in the ground-state energy.
The same cancellation between the direct energy contributions of the purely imaginary roots and those induced by changes in the root densities also occurs for ground-state configurations in other boundary-parameter ranges, as shown in Sec.~\ref{sec:boundary-string-energies}.
Thus, Eq.~\eqref{eq:ground-state-energy} retains the same form despite changes in the ground-state boundary-string configuration.
The extensive $O(N)$ term agrees with the bulk ground-state energy of the periodic chain obtained by Martins~\cite{martins}, after accounting for the energy normalization used here.

Table~\ref{tab:ground-energy-boundary-orientation} compares this analytical prediction with DMRG energies for five boundary orientations at fixed $C$ and $\tilde C$, illustrating their close agreement and the small orientation dependence remaining at finite size.

\begin{table}[htbp]
\centering

\caption{ Comparison of analytical and DMRG ground-state energies of the open chain for different boundary-parameter sets at fixed boundary magnitudes.
Here $N=100$ and $\tilde c_1=\tilde c_2=0.25$.
The five pairs $(c_1,c_2)$ satisfy $C=2(c_1^2+c_2^2)=6.25$, while $\tilde C=0.25$, corresponding to $s_{\rm L}=0.4$ and $s_{\rm R}=2$.
Equation~\eqref{eq:ground-state-energy} depends on the boundary parameters only through $C$ and $\tilde C$ and therefore gives the same analytical energy $E_g^{\rm th}$ for all five sets.
The DMRG energies $E_g^{\rm DMRG}$ differ from this prediction by less than $10^{-2}$, with a spread of $2.17\times10^{-3}$ among the five sets, showing a weak residual dependence on the boundary orientation at finite size.}
\label{tab:ground-energy-boundary-orientation}
\begin{tabular}{cccc}
\toprule
$c_1$ & $c_2$ & $E_g^{\rm th}$ & $E_g^{\rm DMRG}$ \\
\midrule
$0.25$ & $1.75$ & $-100.06101$ & $-100.06874$ \\
$0.85$ & $1.55$ & $-100.06101$ & $-100.07048$ \\
$1.25$ & $1.25$ & $-100.06101$ & $-100.07091$ \\
$1.55$ & $0.85$ & $-100.06101$ & $-100.07048$ \\
$1.75$ & $0.25$ & $-100.06101$ & $-100.06874$ \\
\bottomrule
\end{tabular}

\end{table}

\begin{figure}[htbp]
\centering
\includegraphics[width=0.6\textwidth]{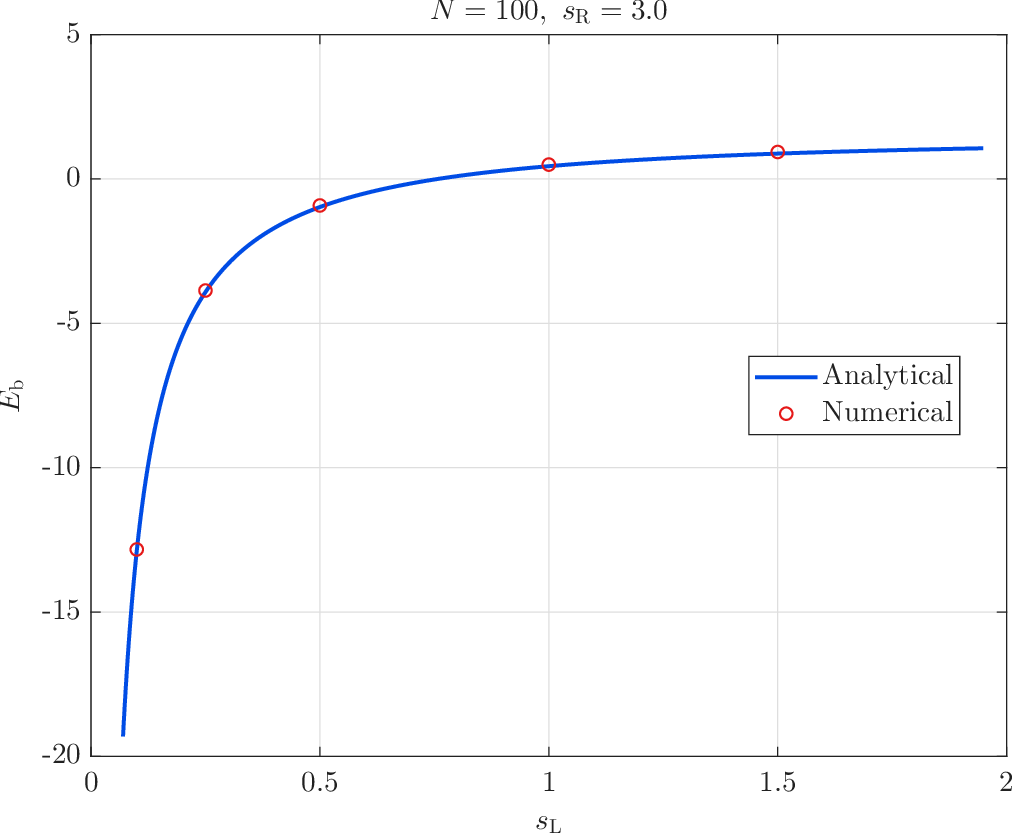}
\caption{Surface energy as a function of $s_{\rm L}$ at $s_{\rm R}=3$ and $N=100$.
The boundary parameters are $c_1=c_2=\tfrac{1}{2s_{\rm L}}$ and $\tilde c_1=\tilde c_2=\tfrac16$.
The blue solid line shows the analytical result $E_b$, and the red circles show the numerical result $E_b^{\rm num}$.}
\label{fig:surface-energy}
\end{figure}

The surface energy is defined as the difference between the ground-state energies of the open and periodic chains with the same number of sites \cite{gaudin1971},
\begin{equation}
E_b=E_g-E_p,
\label{eq:surface-energy-definition}
\end{equation}
where $E_p$ denotes the ground-state energy of the periodic chain with the same bulk Hamiltonian and energy normalization.
Subtracting the bulk contribution identified above from Eq.~\eqref{eq:ground-state-energy}, we obtain the surface energy in the thermodynamic limit,
\begin{equation}
\begin{aligned}
E_b
={}&
-3-2\ln 2+\frac{(9+\sqrt{3})\pi}{9}
+\sqrt{3}\ln(2+\sqrt{3})
\\
&-\frac{3}{2}\sqrt C-\frac{3}{2}\sqrt{\tilde C}
+\frac{3}{2}\int_{-\infty}^{\infty}
\frac{
e^{-|\omega|/\sqrt C}
+e^{-|\omega|/\sqrt{\tilde C}}
}{
1+e^{3|\omega|}
}\,d\omega.
\end{aligned}
\label{eq:surface-energy}
\end{equation}
For comparison, we compute the finite-size surface energy
\begin{equation}
E_b^{\rm num}
=
E_g^{\rm DMRG}-E_p^{\rm BA},
\label{eq:numerical-surface-energy}
\end{equation}
where $E_g^{\rm DMRG}$ is the open-chain ground-state energy obtained by DMRG, and $E_p^{\rm BA}$ is the periodic-chain ground-state energy calculated by numerically solving the Bethe ansatz equations in Ref.~\cite{martins}.
Figure~\ref{fig:surface-energy} compares $E_b^{\rm num}$ with the analytical surface energy $E_b$ in Eq.~\eqref{eq:surface-energy} as $s_{\rm L}$ varies, with $s_{\rm R}=3$ and $N=100$.

\begin{figure}[htbp]
\centering
\includegraphics[width=0.6\textwidth]{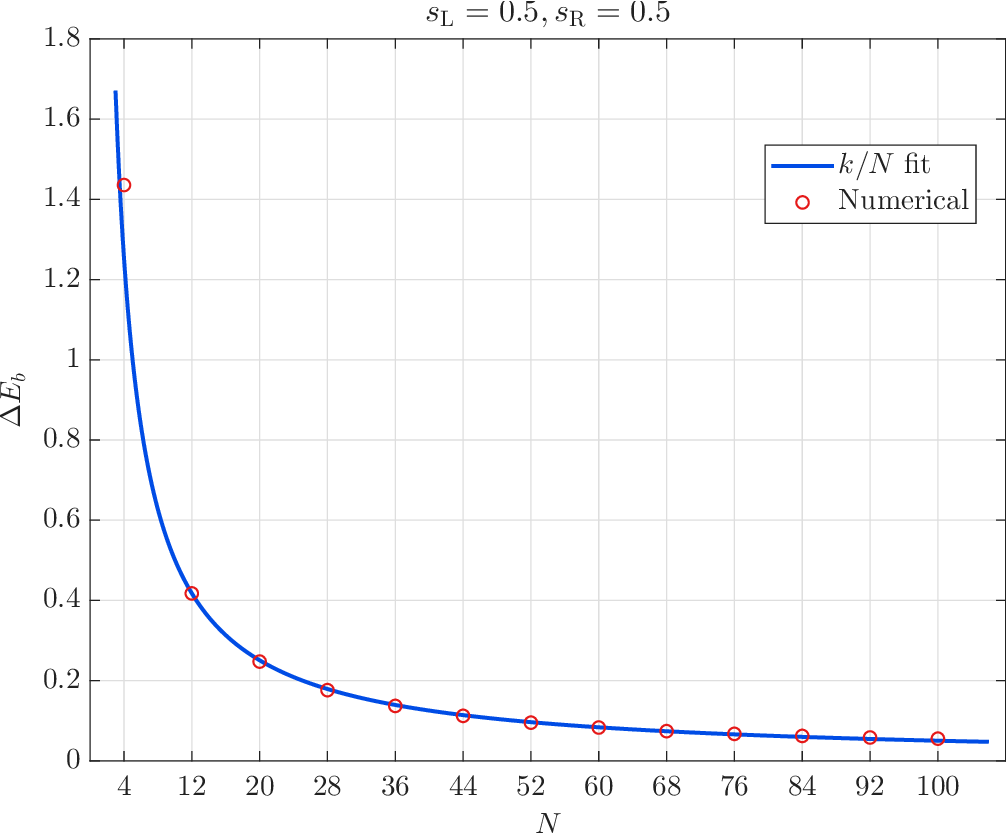}
\caption{Finite-size deviation $\Delta E_b$ versus $N$ at fixed boundary parameters $c_1=c_2=\tilde c_1=\tilde c_2=1$, corresponding to $s_{\rm L}=s_{\rm R}=0.5$.
The red circles show the numerical results, and the blue solid line is a $\tfrac{k}{N}$ fit with $k=5.01288$.}
\label{fig:surface-energy-scaling}
\end{figure}

To analyze the finite-size corrections, we fix the boundary parameters and vary the chain length $N$.
We define the deviation from the analytical surface energy as
\begin{equation}
\Delta E_b=E_b^{\rm num}-E_b.
\label{eq:surface-energy-deviation}
\end{equation}
For the boundary parameters used in Fig.~\ref{fig:surface-energy-scaling}, the numerical results closely follow $\Delta E_b=\tfrac{k}{N}$ with $k=5.01288$, supporting an $O(\tfrac{1}{N})$ finite-size correction.

We have thus determined the ground-state root densities and energy in the thermodynamic limit. The calculation shows that the direct energy contributions of the purely imaginary roots in Eq.~\eqref{eq:boundary-root-range} are cancelled by those arising from the induced changes in the root densities. Some of these roots have positions determined by the boundary parameters and form boundary strings. 
We now investigate the boundary-string positions analytically, first for the ground state and then for admissible configurations across the spectrum.

\section{Boundary-String Configurations}
\label{sec:boundary-strings}
\setcounter{equation}{0}
We now turn to the boundary strings, whose positions and allowed combinations depend on the boundary parameters.
We first determine their ground-state configurations and then extend the analysis to other states, deriving a selection rule for the allowed boundary-string combinations.
We then evaluate the energy contributions associated with these configurations.

\subsection{Ground-State Boundary Strings}
\label{sec:ground-boundary-strings}
We first analytically determine the boundary-string positions for the ground state discussed in Sec.~\ref{sec:ground-state}.
Following the regularity analysis of Ref.~\cite{qiao2}, we examine the Fourier-transformed root densities in Eqs.~\eqref{eq:rho-y} and \eqref{eq:rho-w}.
A well-defined thermodynamic root distribution requires the root densities to remain finite in the limit $|\omega|\rightarrow\infty$.
Since $q=e^{-|\omega|/2}$, this condition requires all divergent negative powers of $q$ to cancel as $q\rightarrow0$.

Since $s_{\rm R}>\tfrac32$ in the parameter regime considered here, the right-boundary terms $q^{2s_{\rm R}-2}$ in Eq.~\eqref{eq:rho-y} and $q^{2s_{\rm R}-3}$ in Eq.~\eqref{eq:rho-w} remain finite as $q\rightarrow0$.
We can thus determine the boundary strings by decreasing $s_{\rm L}$ from the nondivergent region and identifying the values at which new divergent terms appear.

For $\tilde{\rho}_y(\omega)$, the left-boundary term $q^{2s_{\rm L}-2}$ remains finite for $s_{\rm L}>1$.
As $s_{\rm L}$ decreases below $1$, this term becomes divergent,
\begin{equation}
0<s_{\rm L}<1.
\end{equation}
The divergence must be cancelled by the contribution associated with the finite purely imaginary roots $\pm i\alpha$.
Matching the powers of $q$ gives
\begin{equation}
2\alpha-5=2s_{\rm L}-2,
\end{equation}
and hence
\begin{equation}
\alpha=\tfrac{3}{2}+s_{\rm L},
\qquad
0<s_{\rm L}<1.
\label{eq:alpha-locking}
\end{equation}

For $\tilde{\rho}_w(\omega)$, the left-boundary term $q^{2s_{\rm L}-3}$ remains finite for $s_{\rm L}>\tfrac{3}{2}$.
When $s_{\rm L}$ decreases below $\tfrac{3}{2}$, this term becomes divergent,
\begin{equation}
0<s_{\rm L}<\tfrac{3}{2}.
\end{equation}
This divergence is cancelled by the contribution associated with $\pm i\beta_1$.
Matching the powers of $q$ gives
\begin{equation}
2\beta_1-5=2s_{\rm L}-3,
\end{equation}
and hence
\begin{equation}
\beta_1=1+s_{\rm L},
\qquad
0<s_{\rm L}<\tfrac{3}{2}.
\label{eq:beta1-locking}
\end{equation}
With $\beta_1$ fixed to this boundary-string value, the divergence proportional to $q^{2s_{\rm L}-3}$ is removed.
For $\tfrac{1}{2}<s_{\rm L}<\tfrac{3}{2}$, the remaining terms are still finite.
As $s_{\rm L}$ is decreased further below $\tfrac{1}{2}$, however, the remaining term proportional to $q^{2s_{\rm L}-1}$ becomes divergent,
\begin{equation}
0<s_{\rm L}<\tfrac{1}{2}.
\end{equation}
A second boundary string $\pm i\beta_2$ is then required to cancel this new divergence.
Matching the corresponding powers gives
\begin{equation}
2\beta_2-5=2s_{\rm L}-1,
\end{equation}
and consequently
\begin{equation}
\beta_2=2+s_{\rm L},
\qquad
0<s_{\rm L}<\tfrac{1}{2}.
\label{eq:beta2-locking}
\end{equation}
These analytical results reproduce the ground-state boundary-string configurations summarized in Tables~\ref{tab:Lambda+} and \ref{tab:Lambdabar+}.

\subsection{Boundary-String Excitation and Selection Rule}
\label{sec:boundary-string-selection-rule}
It is important to note that the boundary-string positions derived above were obtained not from regularity alone, but under constraints inferred from the numerical ground-state configurations: for $A>0$ and $s_{\rm R}>\tfrac32$, the configurations in Tables~\ref{tab:Lambda+} and \ref{tab:Lambdabar+} contain one finite purely imaginary root pair in $\Lambda(u)$ and two in $\bar\Lambda(u)$, with their ranges specified in Eq.~\eqref{eq:boundary-root-range}.
If these restrictions on the root numbers and ranges are relaxed, the regularity conditions admit additional boundary-string configurations and constrain their allowed combinations, which correspond to boundary-string excitations.

To determine these configurations, we introduce arbitrary sets of finite purely imaginary crossing pairs
\begin{equation}
\{\pm i \alpha_\mu\}_{\mu=1}^{M_a}\subset\Lambda(u),
\qquad
\{\pm i \beta_\nu\}_{\nu=1}^{M_b}\subset\bar\Lambda(u),
\qquad
\alpha_\mu>0,\quad \beta_\nu>0,
\label{eq:general-finite-pairs}
\end{equation}
where $M_a$ and $M_b$ are arbitrary non-negative integers.
With these finite pairs included, the complete Fourier-transformed root densities take the form
\begin{equation}
\begin{aligned}
\tilde{\rho}_x(\omega)
={}&
\frac{2q}{1+q^2}
+\frac{1}{N}
\Bigg[
-\frac{(q-1)^2}{1+q^2}
\\
&\quad
-\sum_{\frac12<\alpha_\mu<\frac32}
\frac{q^{2-2\alpha_\mu}+q^{2\alpha_\mu}}{1+q^2}
-\sum_{0<\alpha_\mu<\frac12}
\left(q^{-2\alpha_\mu}+q^{2\alpha_\mu}\right)
\\
&\quad
+\sum_{0<\beta_\nu<1}
\frac{q^{1-2\beta_\nu}+q^{1+2\beta_\nu}}{1+q^2}
\Bigg],
\end{aligned}
\label{eq:general-rho-x}
\end{equation}
\begin{equation}
\begin{aligned}
\tilde{\rho}_y(\omega)
={}&
\frac{2q^4}{(1+q^2)(1+q^6)}
\\
&+
\frac{1}{N}
\Bigg[
\frac{
(q-1)(q^6+q^4+2q^3+2q^2+q+1)
}{
(1+q^2)^2(1-q^2+q^4)
}
\\
&\quad
+\frac{q^{2s_{\rm L}-2}+q^{2s_{\rm R}-2}}{1+q^6}
-\sum_{\alpha_\mu>\frac32}q^{2\alpha_\mu-5}
\\
&\quad
+\sum_{\frac12<\alpha_\mu<\frac32}
\frac{
q^{7-2\alpha_\mu}+q^{2\alpha_\mu-1}
\left(1-q^4+q^6\right)
}{
(1+q^2)^2(1-q^2+q^4)
}
\\
&\quad
+\sum_{0<\alpha_\mu<\frac12}
\left(q^{-2\alpha_\mu-1}+q^{2\alpha_\mu-1}\right)
\\
&\quad
-\sum_{0<\beta_\nu<1}
\frac{
\left(q^{-2\beta_\nu}+q^{2\beta_\nu}\right)
\left(1+q^2+q^6\right)
}{
(1+q^2)^2(1-q^2+q^4)
}
\Bigg],
\end{aligned}
\label{eq:general-rho-y}
\end{equation}
and
\begin{equation}
\begin{aligned}
\tilde{\rho}_w(\omega)
={}&
\frac{2q^3}{1+q^6}
\\
&+
\frac{1}{N}
\Bigg[
\frac{
(q-1)(q+1)(q^2+q+1)
}{
(1+q^2)(1-q^2+q^4)
}
\\
&\quad
+\frac{q^{2s_{\rm L}-3}+q^{2s_{\rm R}-3}}
{1-q^2+q^4}
\\
&\quad
+\sum_{\frac12<\alpha_\mu<\frac32}
\frac{q^{2\alpha_\mu-4}-q^{-2\alpha_\mu}}{1+q^6}
-\sum_{\beta_\nu>1}q^{2\beta_\nu-5}
\\
&\quad
-\sum_{0<\beta_\nu<1}
\frac{q^{1-2\beta_\nu}+q^{1+2\beta_\nu}}{1+q^6}
\Bigg].
\end{aligned}
\label{eq:general-rho-w}
\end{equation}

The two boundary parameters $s_{\rm L}$ and $s_{\rm R}$ enter Eqs.~\eqref{eq:general-rho-y} and \eqref{eq:general-rho-w} in exactly the same form.
It is therefore sufficient to analyze the boundary strings generated by one boundary, since the result for the other boundary follows by exchanging the roles of the left and right boundaries.
We focus on the left boundary and take $s_{\rm R}>\tfrac{3}{2}$, for which the right-boundary terms contain no negative powers of $q$ as $q\rightarrow0$.

To analyze regularity as $q\to0$, we collect below the terms in $N\tilde{\rho}_\kappa(\omega)$ that can become divergent in the parameter ranges considered, and denote their sums by $\mathcal{B}_\kappa(q)$, with $\kappa=x,y,w$.
Accordingly, the potentially divergent parts of the three root densities are
\begin{equation}
\mathcal{B}_x(q)
=
-\sum_{1<\alpha_\mu<\frac32}q^{2-2\alpha_\mu}
-\sum_{0<\alpha_\mu<\frac12}q^{-2\alpha_\mu}
+\sum_{\frac12<\beta_\nu<1}q^{1-2\beta_\nu},
\label{eq:Bx}
\end{equation}
\begin{equation}
\begin{aligned}
\mathcal{B}_y(q)
={}&
q^{2s_{\rm L}-2}
-\sum_{\frac32<\alpha_\mu<\frac52}q^{2\alpha_\mu-5}
+\sum_{0<\alpha_\mu<\frac12}
\left(
q^{-2\alpha_\mu-1}+q^{2\alpha_\mu-1}
\right)
-\sum_{0<\beta_\nu<1}q^{-2\beta_\nu},
\end{aligned}
\label{eq:By}
\end{equation}
and
\begin{equation}
\begin{aligned}
\mathcal{B}_w(q)
={}&
q^{2s_{\rm L}-3}
+q^{2s_{\rm L}-1}
+\sum_{\frac12<\alpha_\mu<\frac32}
\left(
q^{2\alpha_\mu-4}-q^{-2\alpha_\mu}
\right)
\\
&-
\sum_{1<\beta_\nu<\frac52}q^{2\beta_\nu-5}
-\sum_{\frac12<\beta_\nu<1}q^{1-2\beta_\nu}.
\end{aligned}
\label{eq:Bw}
\end{equation}

Matching the negative powers in Eqs.~\eqref{eq:Bx}--\eqref{eq:Bw}, and continuing the cancellation whenever an introduced finite pair generates additional divergent terms, we obtain all boundary-string positions compatible with a complete regular configuration.
The resulting possible positions are summarized in Table~\ref{tab:boundary-string-positions}.
\begin{table}[htbp]
\centering
\caption{Possible boundary-string half-distances associated with the left boundary in different parameter regions.
For $0<s_{\rm L}<\tfrac32$, left-boundary strings must occur in at least one of $\Lambda(u)$ and $\bar\Lambda(u)$, but need not occur in both; for $s_{\rm L}>\tfrac32$, no $s_{\rm L}$-locked boundary strings occur.
Admissible combinations of the listed positions must cancel all divergent terms in $\mathcal{B}_x(q)$, $\mathcal{B}_y(q)$, and $\mathcal{B}_w(q)$.}
\label{tab:boundary-string-positions}
\begin{tabular}{ccc}
\toprule
Range of $s_{\rm L}$ & $\Lambda(u)$ & $\bar\Lambda(u)$ \\
\midrule
$0<s_{\rm L}<\tfrac12$ &
$\begin{array}{@{}c@{}}
\tfrac12-s_{\rm L},\quad \tfrac12+s_{\rm L},\quad \tfrac32-s_{\rm L}\\
\tfrac32+s_{\rm L},\quad \tfrac52-s_{\rm L}
\end{array}$ &
$\begin{array}{@{}c@{}}
s_{\rm L},\quad 1-s_{\rm L},\quad 1+s_{\rm L}\\
2-s_{\rm L},\quad 2+s_{\rm L}
\end{array}$\\\midrule
$\tfrac12<s_{\rm L}<1$ &
$\begin{array}{@{}c@{}}
\tfrac32-s_{\rm L},\quad \tfrac32+s_{\rm L}
\end{array}$ &
$\begin{array}{@{}c@{}}
1-s_{\rm L},\quad 1+s_{\rm L},\quad 2-s_{\rm L}
\end{array}$\\\midrule
$1<s_{\rm L}<\tfrac32$ &
$\begin{array}{@{}c@{}}
\tfrac32-s_{\rm L},\quad \tfrac72-s_{\rm L}
\end{array}$ &
$\begin{array}{@{}c@{}}
s_{\rm L}-1,\quad 2-s_{\rm L},\quad 1+s_{\rm L}
\end{array}$\\\midrule
$s_{\rm L}>\tfrac32$ &
No $s_{\rm L}$-locked position &
No $s_{\rm L}$-locked position\\
\bottomrule
\end{tabular}
\end{table}

The positions listed in Table~\ref{tab:boundary-string-positions} should be understood as the union of all boundary-string positions that can occur in complete regular configurations, rather than as independent solutions.
A finite pair introduced to cancel one divergent term may simultaneously generate additional negative powers in another density channel.

This requirement gives the following \textbf{boundary-string selection rule}: \textit{An admissible combination must simultaneously cancel all divergent terms in $\mathcal{B}_x(q)$, $\mathcal{B}_y(q)$, and $\mathcal{B}_w(q)$ as $q\to0$.}

Thus, the boundary-string positions listed in Table~\ref{tab:boundary-string-positions} cannot be combined independently.
Each realized boundary-string configuration has a corresponding lowest-energy state in the spectrum, which we call its base state following Ref.~\cite{zhakenov2025}.

\begin{figure}[htbp]
\centering
\begin{tabular}{cc}
\includegraphics[width=0.47\textwidth]{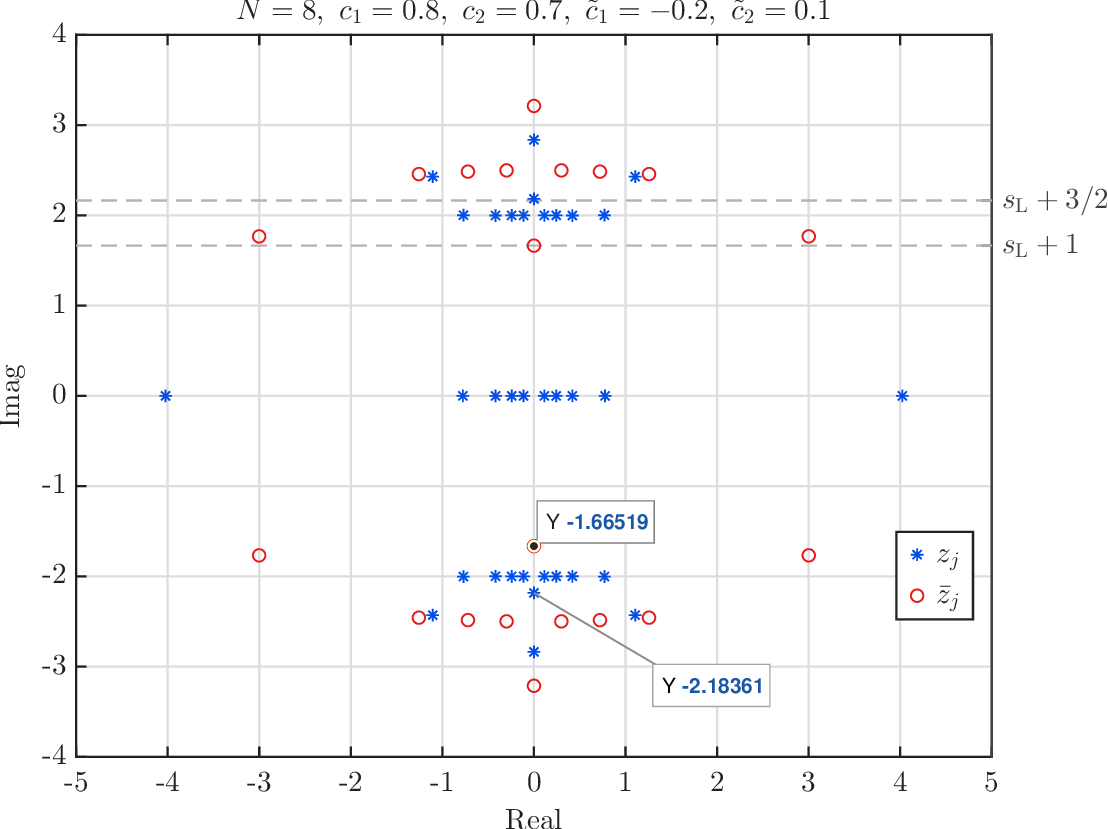} &
\includegraphics[width=0.46\textwidth]{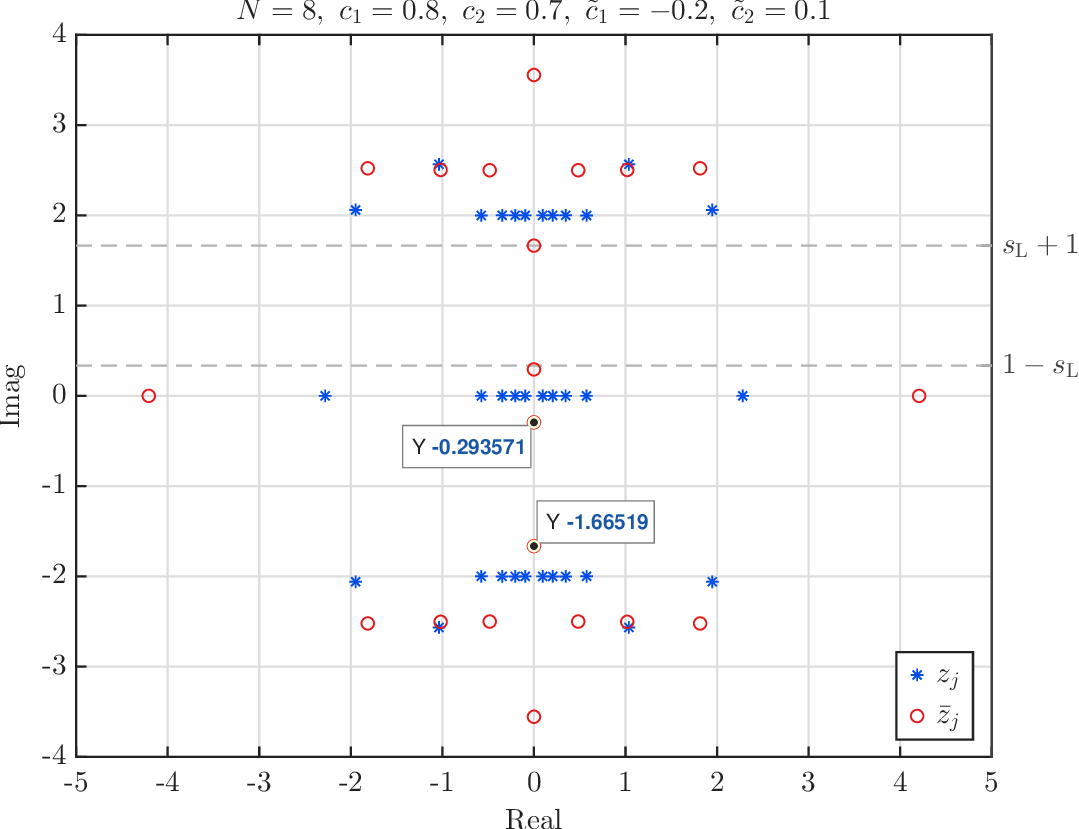} \\
(a) Ground state & (b) Base state
\end{tabular}
\caption{Zero-root configurations of (a) the ground state and (b) the base state of the boundary-string configuration in Eq.~\eqref{eq:boundary-string-excited-example}, corresponding to the twentieth excited state.
The boundary parameters are the same as in Fig.~\ref{fig:ground-representatives}(d), but the chain is shorter.
Blue stars and red circles denote $z_j$ of $\Lambda(u)$ and $\bar z_j$ of $\bar\Lambda(u)$, respectively.}
\label{fig:boundary-string-excited}
\end{figure}

Besides the ground-state boundary-string configuration discussed above, other admissible combinations also occur in the excited-state spectrum.
For example, for $\tfrac12<s_{\rm L}<1$, we find a configuration with no boundary strings in $\Lambda(u)$ and two boundary-string pairs in $\bar\Lambda(u)$:
\begin{equation}
\bar\Lambda(u):\quad
\beta_1=1-s_{\rm L},\qquad
\beta_2=1+s_{\rm L}.
\label{eq:boundary-string-excited-example}
\end{equation}
This combination satisfies the regularity conditions of Eqs.~\eqref{eq:Bx}--\eqref{eq:Bw}.
Fig.~\ref{fig:boundary-string-excited} compares the ground state with the twentieth excited state, which is the base state of this boundary-string configuration.
Both states are calculated with the same boundary parameters as in Fig.~\ref{fig:ground-representatives}(d), but for a shorter chain.

The boundary-string selection rule allows base states with different numbers of boundary strings.
Owing to the zero-root symmetry established in Appendix~\ref{app:zero-root-symmetry}, these changes may cause the root distributions to exhibit the bulk-excitation configurations described in Sec.~\ref{sec:excitations}.
Thus, relative to the ground state, a base state may also contain bulk excitations in addition to its different boundary-string configuration.

\subsection{Boundary-String Excitation Energies}
\label{sec:boundary-string-energies}

To determine the boundary-string excitation energies, we first set aside the boundary-string selection rule and evaluate the contribution of a single purely imaginary root pair.
Following Eq.~\eqref{eq:general-finite-pairs}, $\alpha$ and $\beta$ denote the positive half-distances of the purely imaginary root pairs in $\Lambda(u)$ and $\bar\Lambda(u)$, respectively.
We extract the density terms associated with a single $\alpha$ or $\beta$ pair from Eqs.~\eqref{eq:general-rho-x}--\eqref{eq:general-rho-w} and denote them by $\delta\tilde\rho_\kappa^{(\alpha)}$ and $\delta\tilde\rho_\kappa^{(\beta)}$, respectively, with $\kappa=x,y,w$.
Since these density corrections add linearly, the energy contribution of each pair can be evaluated separately.
	
We denote the contribution of a single purely imaginary root pair to the total energy by $\varepsilon_\alpha(\alpha)$ or $\varepsilon_\beta(\beta)$, including both the energy contribution from the induced changes in the bulk-string densities and the explicit root contribution.
Equation~\eqref{eq:energy-thermodynamic} gives
\begin{equation}
\begin{aligned}
\varepsilon_\alpha(\alpha)
={}&\frac{3N}{2}\int_0^\infty
\left[(q^3+q^7-q)\delta\tilde\rho_x^{(\alpha)}
+(q^8-q^2)\delta\tilde\rho_y^{(\alpha)}\right]d\omega
+\frac{9}{2(\tfrac94-\alpha^2)},\\
\varepsilon_\beta(\beta)
={}&\frac{3N}{2}\int_0^\infty
\left[(q^3+q^7-q)\delta\tilde\rho_x^{(\beta)}
+(q^8-q^2)\delta\tilde\rho_y^{(\beta)}\right]d\omega.
\end{aligned}
\label{eq:finite-pair-energy}
\end{equation}
Substituting the single-pair density corrections from Eqs.~\eqref{eq:general-rho-x} and \eqref{eq:general-rho-y} into Eq.~\eqref{eq:finite-pair-energy}, we obtain
\begin{equation}
\varepsilon_\alpha(\alpha)=
\begin{cases}
0,&0<\alpha<\tfrac12,\\
\varepsilon_{b}(\alpha+\tfrac12),&\tfrac12<\alpha<\tfrac32,\\
0,&\alpha>\tfrac32,
\end{cases}
\qquad
\varepsilon_\beta(\beta)=
\begin{cases}
\varepsilon_{b}(\beta),&0<\beta<1,\\
0,&\beta>1,
\end{cases}
\label{eq:finite-pair-energy-piecewise}
\end{equation}
where
\begin{equation}
\begin{aligned}
\varepsilon_{b}(t)
&=\frac32\int_0^\infty
\frac{\cosh(t\omega)}{\cosh(\omega/2)\cosh(3\omega/2)}\,d\omega\\
&=\frac{\pi}{\sin(\pi t)}
\left[\frac{2}{\sqrt3}\sin\left(\frac{2\pi t}{3}\right)-t\right].
\end{aligned}
\label{eq:boundary-energy-function}
\end{equation}
	
Only root pairs with $\tfrac12<\alpha<\tfrac32$ or $0<\beta<1$ give nonzero energy contributions, all of which are positive.
As shown above, the ground-state boundary roots satisfy $\alpha>\tfrac32$ and $\beta>1$, for which the combined contributions from the density corrections and explicit root terms are zero.
Thus, at fixed boundary parameters, summing the contributions of the root pairs in an admissible configuration gives its boundary-string excitation energy relative to the ground state:
\begin{equation}
\varepsilon_{boundary}
=\sum_{\frac12<\alpha_\mu<\frac32}
\varepsilon_{b}(\alpha_\mu+\tfrac12)
+\sum_{0<\beta_\nu<1}\varepsilon_{b}(\beta_\nu).
\label{eq:boundary-string-energy-increment}
\end{equation}
% If a change in the boundary-string configuration requires accompanying bulk excitations, their energy contributions must also be included to obtain the total excitation energy of the physical state.
	
The ground-state boundary-string configuration has zero contribution and therefore attains the lower bound of the boundary-string energy.
Consequently, no minimum-energy configuration can contain roots with $\tfrac12<\alpha<\tfrac32$ or $0<\beta<1$.
Although a root with $0<\alpha<\tfrac12$ contributes zero by itself, regularity of Eq.~\eqref{eq:Bx} requires
\begin{equation}
\beta=\alpha+\tfrac12\in(\tfrac12,1).
\label{eq:zero-energy-root-companion}
\end{equation}
This required companion root has a positive energy contribution, so the combination cannot attain the lower bound either.
The minimum-energy boundary-string configuration can therefore contain only roots with
\begin{equation}
\alpha>\tfrac32,\qquad \beta>1.
\label{eq:minimum-energy-boundary-ranges}
\end{equation}
In these ranges, the purely imaginary roots satisfying the boundary-string selection rule reproduce the ground-state boundary-string configurations listed in Tables~\ref{tab:Lambda+} and \ref{tab:Lambdabar+}.
The minimum-energy boundary-string configuration is therefore unique and agrees with the numerical observations.

\section{Elementary Excitations}
\label{sec:excitations}
\setcounter{equation}{0}
The zero-root representation provides a unified description of excitations in both diagonal and non-diagonal boundary cases. Such a unified description is difficult to achieve in the conventional Bethe-root representation, where the root structures differ substantially between the two cases.

We now consider bulk excitations that preserve the ground-state boundary-string configuration.
Numerical analysis of the low-energy spectrum reveals a variety of zero-root configurations distinct from that of the ground state.
Despite their different root structures, the analysis below shows that their bulk excitations can be described in terms of three elementary hole excitations, denoted by $X$, $Y$, and $W$.
The $X$- and $Y$-type excitations correspond to holes in the $x$- and $y$-string distributions of $\Lambda(u)$, respectively, while the $W$-type excitation corresponds to holes in the $w$-string distribution of $\bar\Lambda(u)$.

The three elementary hole excitations, however, cannot occur arbitrarily in a physical eigenstate.
The fixed numbers of zero roots, together with zero-root symmetry, constrain the changes in the three string sectors and lead to a bulk-excitation selection rule.
Consequently, the different excitation patterns observed numerically can be understood as different allowed combinations of these three excitations.

To determine the contribution of each excitation, we first temporarily set aside the selection rule and analyze the three excitation sectors separately.
This procedure is justified by the linear structure of the coupled density equations.
The source terms generated by different holes enter the density equations linearly, and the corresponding changes of the root densities can therefore be superposed.
We may thus consider one excitation sector at a time, determine its effect on the three root densities and its excitation energy, and subsequently impose the bulk-excitation selection rule to construct the physical excitation configurations.

For the analytical treatment, we take the $A>0$ ground state obtained above as the reference state.
Apart from the fixed holes at the origin discussed in Sec.~\ref{sec:bulk-excitation-selection-rule}, this reference state contains no holes in the three bulk-string distributions, making it convenient for identifying the density changes induced by each excitation.
For an excitation of type $\eta=X,Y,W$, we denote the induced changes of the three root densities by
\begin{equation}
\delta\rho_x^{(\eta)}(v),\qquad
\delta\rho_y^{(\eta)}(v),\qquad
\delta\rho_w^{(\eta)}(v),
\qquad
\eta=X,Y,W.
\label{eq:excitation-density-variation}
\end{equation}
Correspondingly, their Fourier transforms are denoted by
\begin{equation}
\delta\tilde{\rho}_x^{(\eta)}(\omega),\qquad
\delta\tilde{\rho}_y^{(\eta)}(\omega),\qquad
\delta\tilde{\rho}_w^{(\eta)}(\omega).
\label{eq:excitation-density-variation-Fourier}
\end{equation}
Thus, $\delta\rho_\mu^{(X)}$, $\delta\rho_\mu^{(Y)}$, and $\delta\rho_\mu^{(W)}$ describe the changes of the $\mu$-string density induced by the $X$-, $Y$-, and $W$-type excitations, respectively.
For each excitation, the corresponding hole source is introduced into the coupled density equations, from which the density variations are determined in Fourier space.
Substituting these results into the energy expression then gives the excitation energy relative to the ground state.

In some ranges of the boundary parameters, additional roots are present in the ground-state configuration.
Their positions are not fixed by the boundary parameters and may change in the excited states considered here, while their numbers remain unchanged.
Although these shifts modify the root densities, they neither contribute to the excitation energy, as follows from Eq.~\eqref{eq:finite-pair-energy-piecewise}, nor change the numbers of strings.
We therefore keep these roots implicit and focus on the three elementary hole excitations, omitting the density responses associated with shifts of the additional roots.

For the numerical illustrations of the three excitations, however, we choose boundary parameters in the $A<0$ region, where the ground state already contains a pair of $w$-holes, as listed in Tables~\ref{tab:Lambda-} and \ref{tab:Lambdabar-}.
The selection rule derived below generally requires elementary excitations to occur in combinations.
The pre-existing $w$-holes allow the characteristic root structures of the $X$-, $Y$-, and $W$-type excitations to be displayed individually, making them easier to identify numerically.

We first analyze the three excitations separately and then derive the selection rule that determines their allowed combinations in the physical eigenstates.

\subsection{\texorpdfstring{$W$}{W}-Type Excitation}
\label{sec:W-type Excitation}
\begin{figure}[htbp]
\centering
\includegraphics[width=0.7\textwidth]{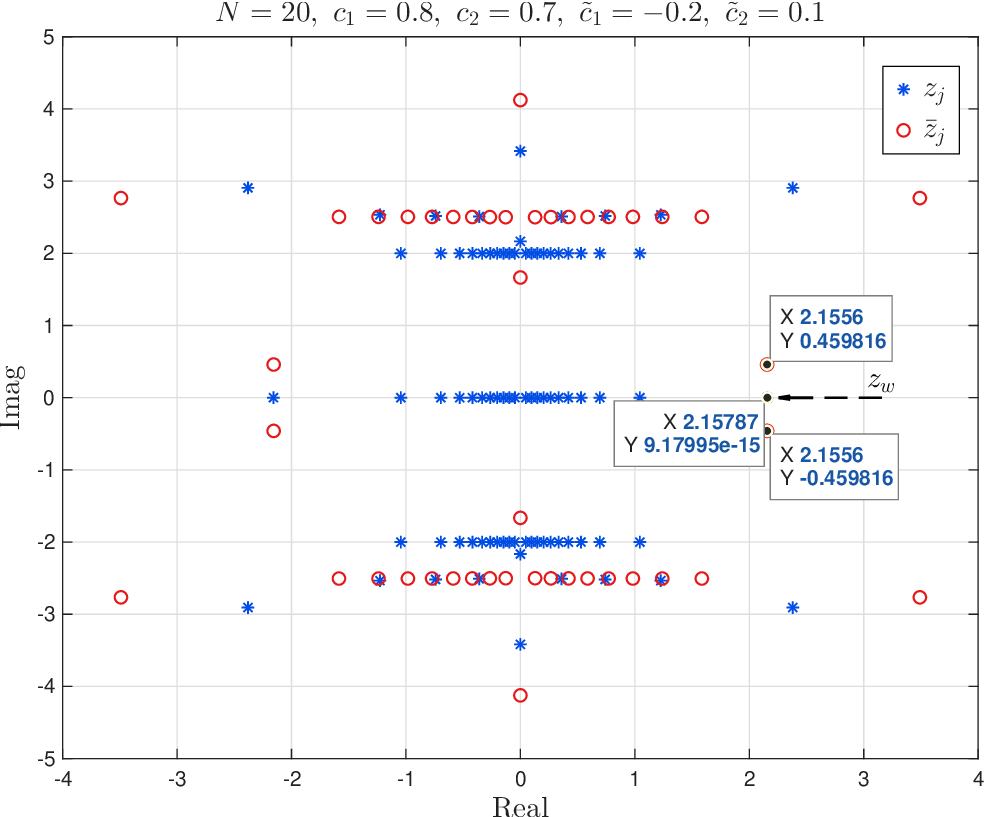}
\caption{Zero-root configuration of the $W$-type excitation in the twelfth excited state, with the corresponding ground state shown in Fig.~\ref{fig:ground-representatives}(d).}
\label{fig:W-excitation}
\end{figure}

We first consider the $W$-type excitation.
As shown in Fig.~\ref{fig:W-excitation}, the numerical zero-root configurations show that this excitation involves finite roots in both $\Lambda(u)$ and $\bar\Lambda(u)$.
For this isolated finite-root configuration, the regularity analysis in Appendix~\ref{app:W-center-locking} fixes the imaginary offsets of the $\bar\Lambda$ quartet to $\pm\tfrac12$ and requires it to share the real center of the $\Lambda$ pair.
Denoting this common center by $z_w$, the $W$-type excitation is characterized by the following zero roots:
\begin{equation}
\Lambda(u):\quad \pm z_w,
\qquad
\bar\Lambda(u):\quad
\pm z_w\pm\tfrac{i}{2},
\qquad
z_w\in\mathbb{R}.
\label{eq:W-root-configuration}
\end{equation}
The remaining zeros retain the corresponding ground-state bulk-string structures. 

To determine the effect of this excitation on the root distributions, we substitute the full root configuration of the $W$-type excited state into the coupled density equations and subtract the corresponding ground-state equations.
The resulting changes of the three root densities are
\begin{equation}
N\delta\tilde{\rho}_x^{(W)}(\omega)
=
0,
\label{eq:W-rho-x}
\end{equation}
\begin{equation}
N\delta\tilde{\rho}_y^{(W)}(\omega)
=
-\frac{2q}{1+q^6}\cos(z_w\omega),
\label{eq:W-rho-y}
\end{equation}
and
\begin{equation}
N\delta\tilde{\rho}_w^{(W)}(\omega)
=
-2\cos(z_w\omega)
-
\frac{2q^2(1-q^4)}{1+q^6}
\cos(z_w\omega).
\label{eq:W-rho-w}
\end{equation}
Denoting the inverse Fourier transform defined in Eq.~\eqref{eq:Fourier-transform} by $\mathcal F^{-1}$, the first term in Eq.~\eqref{eq:W-rho-w} gives
\begin{equation}
\mathcal F^{-1}\!\left[-2\cos(z_w\omega)\right](v)
=
-\delta(v-z_w)-\delta(v+z_w),
\label{eq:W-hole-part}
\end{equation}

showing that one $w$-string is removed at each of the two positions $\pm z_w$ and hence that a pair of holes is created in the $w$-string sea.
The remaining term in Eq.~\eqref{eq:W-rho-w}, together with Eq.~\eqref{eq:W-rho-y}, describes the accompanying smooth redistribution of the $w$- and $y$-string distributions, while Eq.~\eqref{eq:W-rho-x} shows that the $x$-string distribution remains unchanged.
The zero-root configuration in Eq.~\eqref{eq:W-root-configuration} is therefore identified as a $w$-hole excitation.

Using the density variations in Eqs.~\eqref{eq:W-rho-x}--\eqref{eq:W-rho-w} together with the finite roots of $\Lambda(u)$ in Eq.~\eqref{eq:W-root-configuration}, the energy of the $W$-type excited state can be obtained from Eq.~\eqref{eq:energy-thermodynamic}.
Denoting this energy by $E_W$, the excitation energy relative to the ground state is
\begin{equation}
\varepsilon_W(z_w)
=
E_W-E_g
=
\pi\,\operatorname{sech}
\left(\frac{\pi z_w}{3}\right).
\label{eq:W-excitation-energy}
\end{equation}

\subsection{\texorpdfstring{$X$}{X}-Type Excitation}
\begin{figure}[htbp]
\centering
\includegraphics[width=0.7\textwidth]{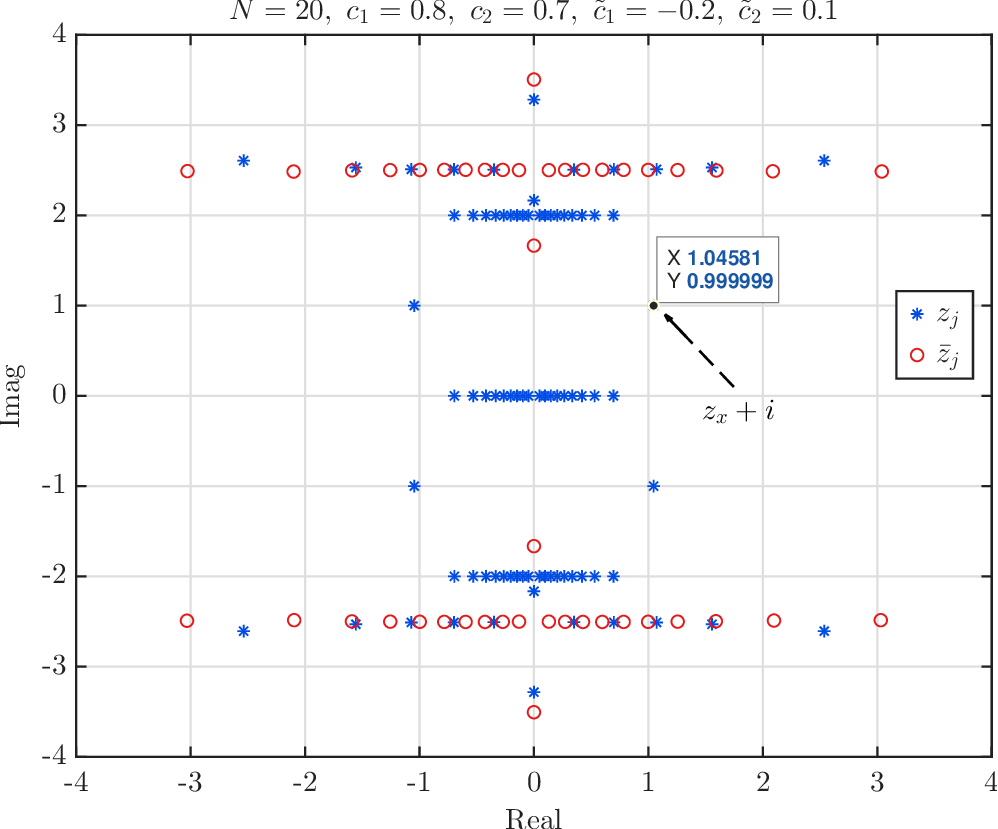}
\caption{Zero-root configuration of the $X$-type excitation in the fifteenth excited state, with the corresponding ground state shown in Fig.~\ref{fig:ground-representatives}(d).}
\label{fig:X-excitation}
\end{figure}

We next consider the $X$-type excitation.
As shown in Fig.~\ref{fig:X-excitation}, the $X$-type excitation is characterized by the following zero roots:
\begin{equation}
\Lambda(u):\quad \pm z_x\pm i,
\qquad
z_x\in\mathbb{R}.
\label{eq:X-root-configuration}
\end{equation}
The other zeros preserve the bulk-string forms of the corresponding ground state.

Inserting the full zero-root configuration of the $X$-type excited state into the coupled density equations and subtracting the ground-state equations, we obtain
\begin{equation}
N\delta\tilde{\rho}_x^{(X)}(\omega)
=
-2\cos(z_x\omega),
\label{eq:X-rho-x}
\end{equation}
\begin{equation}
N\delta\tilde{\rho}_y^{(X)}(\omega)
=
\frac{2q}{1+q^2}\cos(z_x\omega),
\label{eq:X-rho-y}
\end{equation}
and
\begin{equation}
N\delta\tilde{\rho}_w^{(X)}(\omega)
=
0.
\label{eq:X-rho-w}
\end{equation}

The hole structure is contained directly in the $x$-string density.
Taking the inverse Fourier transform of Eq.~\eqref{eq:X-rho-x}, we obtain
\begin{equation}
N\delta\rho_x^{(X)}(v)
=
-\delta(v-z_x)-\delta(v+z_x).
\label{eq:X-rho-x-real}
\end{equation}
Thus, one $x$-string is removed at each of the two positions $\pm z_x$, producing a pair of holes in the $x$-string sea.
At the same time, Eq.~\eqref{eq:X-rho-y} describes the smooth redistribution induced in the $y$-string sea through the coupling of the density equations, whereas Eq.~\eqref{eq:X-rho-w} shows that the $w$-string distribution remains unchanged.
The zero-root configuration in Eq.~\eqref{eq:X-root-configuration} is therefore identified as an $x$-hole excitation.

Using the density variations in Eqs.~\eqref{eq:X-rho-x}--\eqref{eq:X-rho-w} together with the finite roots in Eq.~\eqref{eq:X-root-configuration}, the energy of the $X$-type excited state can be obtained from Eq.~\eqref{eq:energy-thermodynamic}.
Denoting this energy by $E_X$, the excitation energy relative to the ground state is
\begin{equation}
\varepsilon_X(z_x)
=
E_X-E_g
=
3\pi\,\operatorname{sech}(\pi z_x).
\label{eq:X-excitation-energy}
\end{equation}

\subsection{\texorpdfstring{$Y$}{Y}-Type Excitation}
We finally consider the $Y$-type excitation.

\begin{figure}[htbp]
\centering
\includegraphics[width=0.7\textwidth]{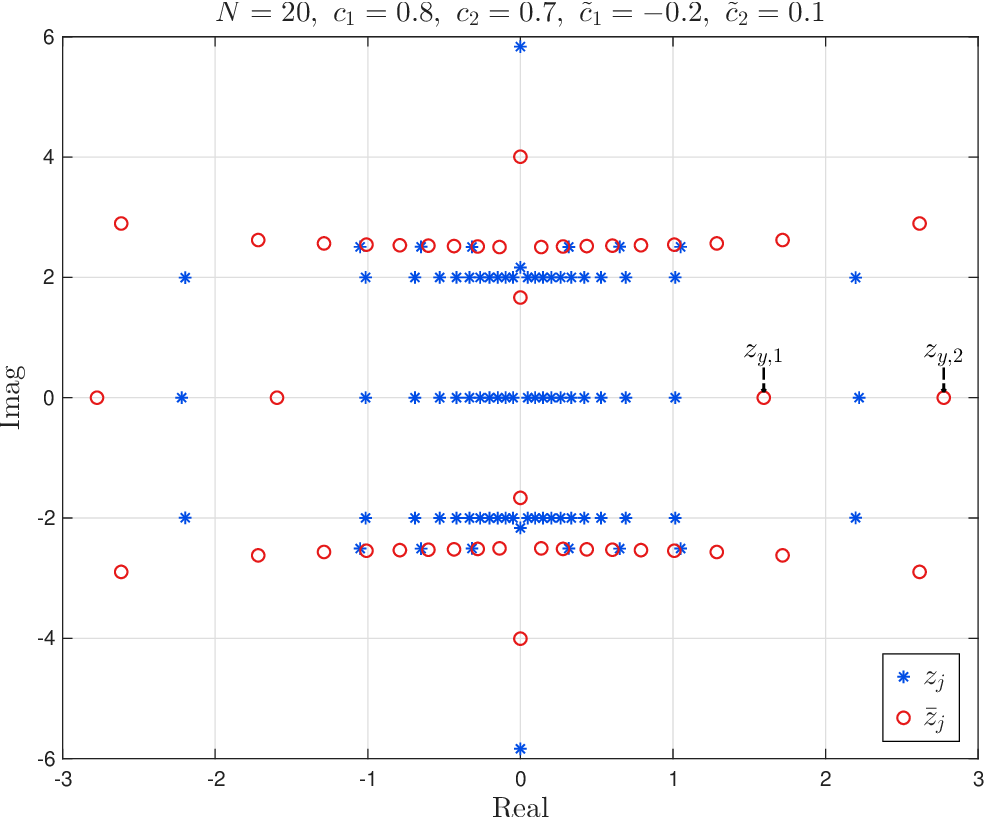}
\caption{Zero-root configuration of the $Y$-type excitation in the twentieth excited state, with the corresponding ground state shown in Fig.~\ref{fig:ground-representatives}(d).}
\label{fig:Y-excitation}
\end{figure}

The state shown in Fig.~\ref{fig:Y-excitation} contains two $Y$-type elementary excitations, characterized by the root pairs $\pm\bar z_{y,1}$ and $\pm\bar z_{y,2}$.
Below, we calculate the density response and energy associated with a single $Y$-type elementary excitation, whose characteristic roots are

\begin{equation}
\bar\Lambda(u):\quad \pm \bar z_y,
\qquad
\bar z_y\in\mathbb{R}.
\label{eq:Y-root-configuration}
\end{equation}

The full zero-root configuration of the $Y$-type excited state similarly yields the following density variations relative to the ground state:
\begin{equation}
N\delta\tilde{\rho}_x^{(Y)}(\omega)
=
\frac{2q}{1+q^2}\cos(\bar z_y\omega),
\label{eq:Y-rho-x}
\end{equation}
\begin{equation}
N\delta\tilde{\rho}_y^{(Y)}(\omega)
=
-2\cos(\bar z_y\omega)
+
\frac{2q^8}{(1+q^2)(1+q^6)}
\cos(\bar z_y\omega),
\label{eq:Y-rho-y}
\end{equation}
and
\begin{equation}
N\delta\tilde{\rho}_w^{(Y)}(\omega)
=
-\frac{2q}{1+q^6}\cos(\bar z_y\omega).
\label{eq:Y-rho-w}
\end{equation}
The inverse Fourier transform of the first term in Eq.~\eqref{eq:Y-rho-y} gives
\begin{equation}
\mathcal F^{-1}\!\left[-2\cos(\bar z_y\omega)\right](v)
=
-\delta(v-\bar z_y)-\delta(v+\bar z_y),
\label{eq:Y-hole-part}
\end{equation}

showing that one $y$-string is removed at each of the two positions $\pm\bar z_y$ and hence that a pair of holes is created in the $y$-string sea.
The remaining term in Eq.~\eqref{eq:Y-rho-y}, together with Eqs.~\eqref{eq:Y-rho-x} and \eqref{eq:Y-rho-w}, describes the accompanying smooth redistribution of the three root distributions.
The zero-root configuration in Eq.~\eqref{eq:Y-root-configuration} is therefore identified as a $y$-hole excitation.

\begin{figure}[htbp]
\centering
\includegraphics[width=0.6\textwidth]{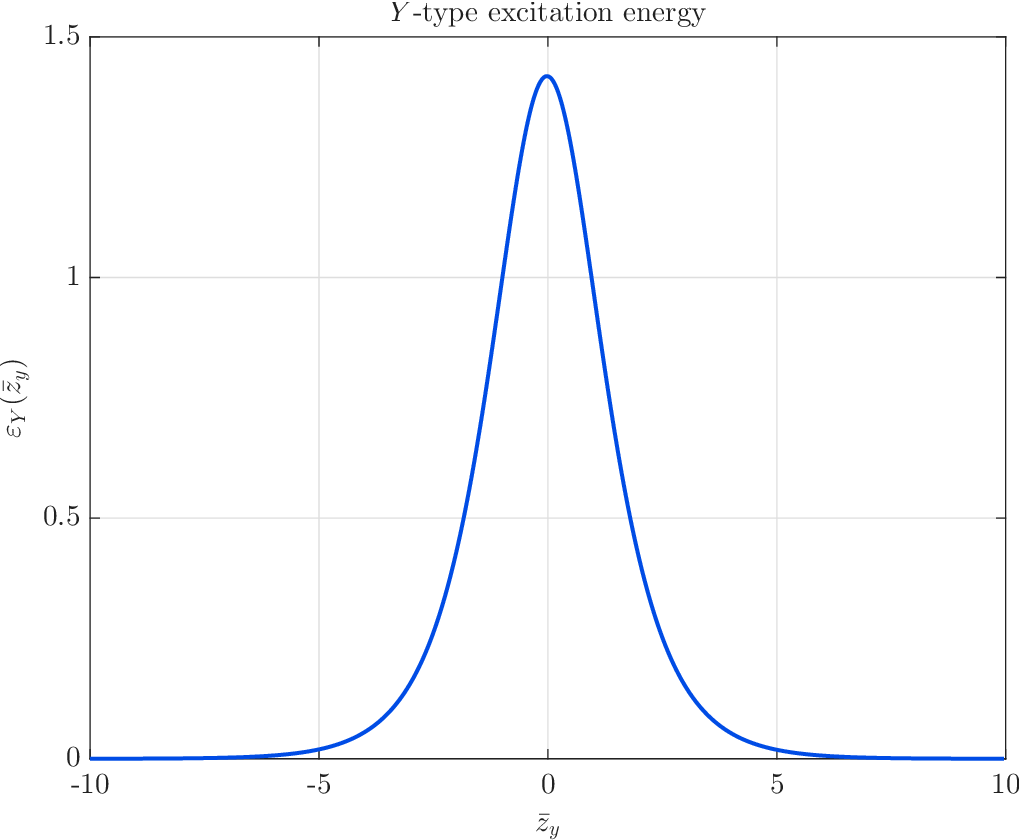}
\caption{$Y$-type excitation energy $\varepsilon_Y$ as a function of $\bar z_y$, given by Eq.~\eqref{eq:Y-excitation-energy}.}
\label{fig:Y-excitation-energy}
\end{figure}

Using the density variations in Eqs.~\eqref{eq:Y-rho-x}--\eqref{eq:Y-rho-w}, the energy of the $Y$-type excited state can be obtained from Eq.~\eqref{eq:energy-thermodynamic}.
Denoting this energy by $E_Y$, the excitation energy relative to the ground state is
\begin{equation}
\begin{aligned}
\varepsilon_Y(\bar z_y)
&=
E_Y-E_g
\\
&=
\frac{3}{2}
\int_0^\infty
\frac{\cos(\bar z_y\omega)}
{\cosh(\omega/2)\cosh(3\omega/2)}
\,d\omega
\\
&=
\frac{\pi}{\sinh(\pi \bar z_y)}
\left[
\frac{2}{\sqrt{3}}
\sinh\left(\frac{2\pi \bar z_y}{3}\right)
-\bar z_y
\right].
\end{aligned}
\label{eq:Y-excitation-energy}
\end{equation}
Figure~\ref{fig:Y-excitation-energy} shows the dependence of $\varepsilon_Y$ on the hole parameter $\bar z_y$.
The energy is an even function of $\bar z_y$, with $\varepsilon_Y(0)=\tfrac{4\pi}{3\sqrt{3}}-1$, and tends to zero as $|\bar z_y|\to\infty$.

\subsection{Bulk-Excitation Selection Rule}
\label{sec:bulk-excitation-selection-rule}

The three elementary excitations derived above cannot be combined arbitrarily in a physical eigenstate.
Their allowed combinations are constrained by the changes in the numbers of the three types of bulk strings.

For $\mu=x,y,w$, we denote the number of occupied $\mu$-strings by $M_\mu$.
According to the definition of the root density,
\begin{equation}
M_\mu
=
N\int_{-\infty}^{\infty}\rho_\mu(v)\,dv
=
N\tilde{\rho}_\mu(0).
\label{eq:string-number}
\end{equation}
For a given reference state, the change in the number of $\mu$-strings is therefore
\begin{equation}
\Delta M_\mu
=
M_\mu-M_\mu^{(\mathrm{ref})}
=
N\delta\tilde{\rho}_\mu(0).
\label{eq:string-number-change}
\end{equation}

Taking the zero-frequency limit of the density variations obtained above, Eqs.~\eqref{eq:X-rho-x}--\eqref{eq:X-rho-w} give
\begin{equation}
\Delta M_x^{(X)}=-2,
\qquad
\Delta M_y^{(X)}=1,
\qquad
\Delta M_w^{(X)}=0,
\label{eq:X-string-number-change}
\end{equation}
for the $X$-type excitation.
Similarly, Eqs.~\eqref{eq:Y-rho-x}--\eqref{eq:Y-rho-w} give
\begin{equation}
\Delta M_x^{(Y)}=1,
\qquad
\Delta M_y^{(Y)}=-\tfrac{3}{2},
\qquad
\Delta M_w^{(Y)}=-1,
\label{eq:Y-string-number-change}
\end{equation}
for the $Y$-type excitation, while Eqs.~\eqref{eq:W-rho-x}--\eqref{eq:W-rho-w} give
\begin{equation}
\Delta M_x^{(W)}=0,
\qquad
\Delta M_y^{(W)}=-1,
\qquad
\Delta M_w^{(W)}=-2,
\label{eq:W-string-number-change}
\end{equation}
for the $W$-type excitation.

Relative to a chosen reference state, we denote the changes in the numbers of $X$-, $Y$-, and $W$-type holes by
\begin{equation}
n_X,n_Y,n_W\in\mathbb{Z}.
\label{eq:excitation-numbers}
\end{equation}
Each unit of $n_X$, $n_Y$, or $n_W$ corresponds to a pair of holes at opposite centers.
Positive and negative values indicate the addition and removal of the corresponding holes, respectively.
By linear superposition, the total changes in the three string numbers are
\begin{equation}
\Delta M_x
=
-2n_X+n_Y,
\label{eq:total-Mx-change}
\end{equation}
\begin{equation}
\Delta M_y
=
n_X-\tfrac{3}{2}n_Y-n_W,
\label{eq:total-My-change}
\end{equation}
and
\begin{equation}
\Delta M_w
=
-n_Y-2n_W.
\label{eq:total-Mw-change}
\end{equation}

The ground-state densities in Eqs.~\eqref{eq:rho-x}--\eqref{eq:rho-w} each contain a constant Fourier term $-1/N$, which gives
\begin{equation}
\mathcal F^{-1}\!\left[-\frac{1}{N}\right](v)
=
-\frac{1}{N}\delta(v).
\label{eq:zero-center-holes}
\end{equation}
This term represents a hole at the origin in each string distribution.
The three excitations considered above preserve these holes, so the excited-state configurations still contain no zero-center strings.
Crossing symmetry and Hermitian conjugation then imply that the $x$-, $y$-, and $w$-strings occur in pairs with opposite nonzero real centers, as shown in Appendix~\ref{app:zero-root-symmetry}.
Consequently, the number of strings of each type is even, and the changes in their occupation numbers satisfy
\begin{equation}
\Delta M_\mu\in2\mathbb{Z},
\qquad \mu\in\{x,y,w\}.
\label{eq:string-number-parity}
\end{equation}
Applying this condition to Eqs.~\eqref{eq:total-Mx-change}--\eqref{eq:total-Mw-change}, we obtain the \textbf{bulk-excitation selection rule}
\begin{equation}
n_X+\tfrac{1}{2}n_Y+n_W
\equiv0
\pmod{2}.
\label{eq:excitation-selection-rule}
\end{equation}
The bulk-excitation selection rule in Eq.~\eqref{eq:excitation-selection-rule} requires $n_Y$ to be even, analogous to the pairwise occurrence of spinons in the periodic spin-$\tfrac12$ XXZ antiferromagnet with an even number of sites.

For $A<0$, the ground state in Fig.~\ref{fig:ground-representatives}(d) already contains a pair of holes in the $w$-string distribution.
Using this or another hole-containing base state as the reference, the above analysis still yields the bulk-excitation selection rule in Eq.~\eqref{eq:excitation-selection-rule}.
At the same chain length and boundary parameters, the excited state in Fig.~\ref{fig:X-excitation} has
\begin{equation}
n_X=1,\qquad
n_Y=0,\qquad
n_W=-1,
\label{eq:Anegative-X-example}
\end{equation}
relative to this ground state, satisfying the bulk-excitation selection rule in Eq.~\eqref{eq:excitation-selection-rule}.
Here $n_W=-1$ corresponds to the removal of the pre-existing $w$-holes in Fig.~\ref{fig:ground-representatives}(d), leaving a root configuration containing only the $X$-type excitation.

\subsection{Numerical Verification of the Bulk-Excitation Selection Rule}
\label{sec:numerical-excitation-selection-rule}
We now test the selection rule for $A>0$, using the hole-free ground state in Fig.~\ref{fig:ground-representatives}(b) as the reference and keeping the chain length and boundary parameters fixed.

\begin{figure}[htbp]
\centering
\includegraphics[width=0.7\textwidth]{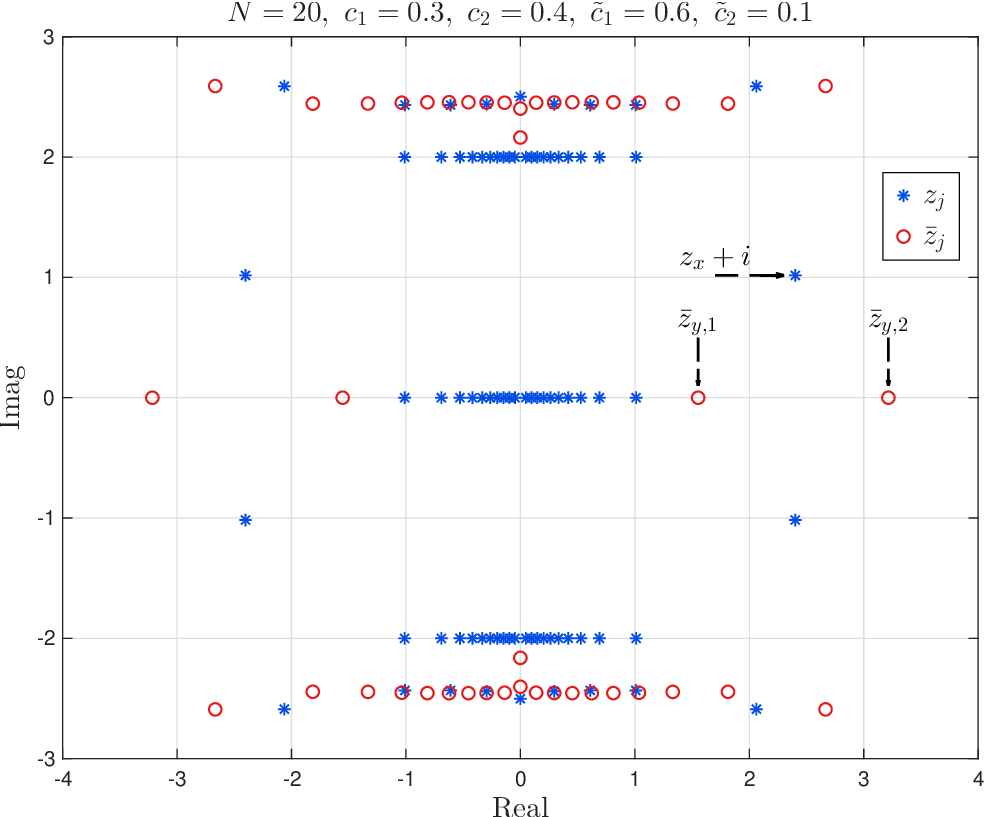}
\caption{Zero-root configuration of the $X+2Y$ excitation in the sixteenth excited state, with the corresponding ground state shown in Fig.~\ref{fig:ground-representatives}(b).}
\label{fig:X2Y-excitation}
\end{figure}

For the sixteenth excited state in Fig.~\ref{fig:X2Y-excitation}, we find
\begin{equation}
\Delta M_x=0,
\qquad
\Delta M_y=-2,
\qquad
\Delta M_w=-2,
\label{eq:X2Y-string-number-change}
\end{equation}
for which
\begin{equation}
n_X=1,\qquad
n_Y=2,\qquad
n_W=0.
\label{eq:X2Y-combination}
\end{equation}
This corresponds to an $X+2Y$ excitation, whose characteristic finite-root configuration is
\begin{equation}
\begin{aligned}
\Lambda(u):\quad&
\pm z_x\pm i,
\\
\bar{\Lambda}(u):\quad&
\pm\bar z_{y,1},\qquad \pm\bar z_{y,2}.
\end{aligned}
\label{eq:X2Y-root-configuration}
\end{equation}
Here $z_x,\bar z_{y,1},\bar z_{y,2}\in\mathbb{R}$.
These occupation-number changes are all even, and $n_X+\tfrac12n_Y+n_W=2$, so this state satisfies the bulk-excitation selection rule.

\begin{figure}[htbp]
\centering
\includegraphics[width=0.9\textwidth]{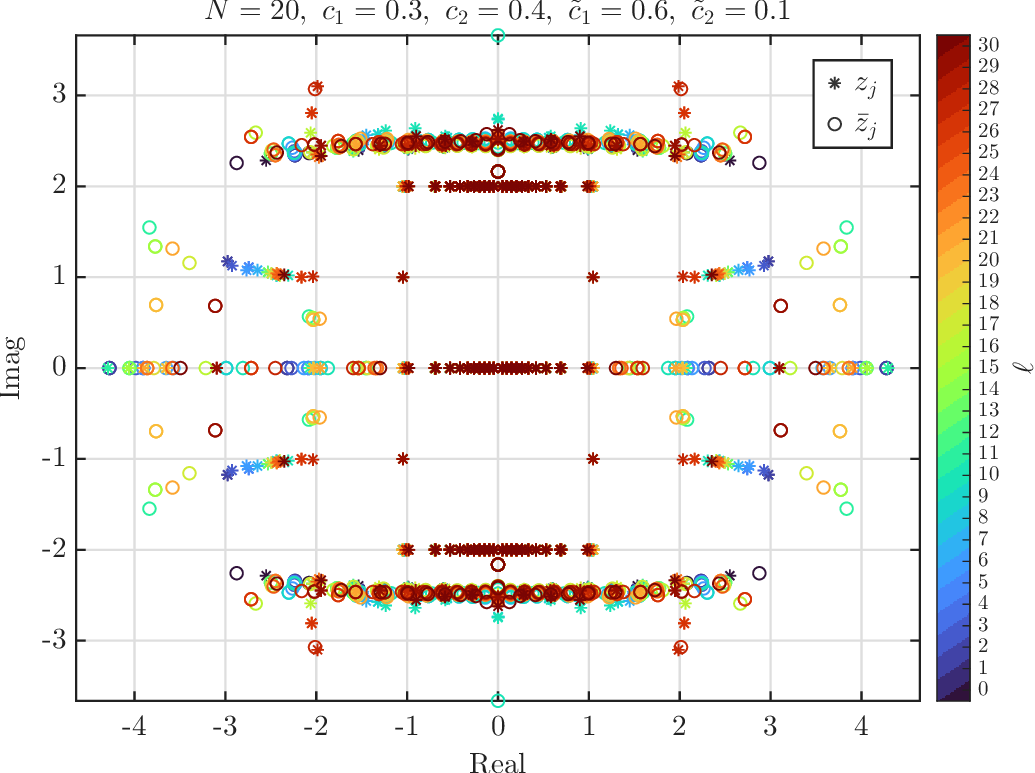}
\caption{Superposed zero-root distributions of the ground state and the first 30 excited states at the same chain length and boundary parameters as in Fig.~\ref{fig:ground-representatives}(b).
Stars and circles denote $z_j$ and $\bar z_j$, respectively; color indicates the state index $\ell$, with $\ell=0$ denoting the ground state.}
\label{fig:excited-state-overlay}
\end{figure}

Figure~\ref{fig:excited-state-overlay} extends this comparison to the ground state and the first 30 excited states.
The corresponding excitation configurations are listed in Table~\ref{tab:numerical-excitation-selection-rule} of Appendix~\ref{app:numerical-excitation-counts} and all satisfy the bulk-excitation selection rule in Eq.~\eqref{eq:excitation-selection-rule}.

\section{Conclusions}
\label{sec:discussion}
\setcounter{equation}{0}

We have investigated the ground-state root structures and low-energy excitations of the open $C_2^{(1)}$ chain with non-diagonal boundaries by combining BA-DMRG calculations with the zero-root method.
Applying the BA-DMRG approach of Ref.~\cite{ji} to both the fundamental and fused transfer matrices yields two complementary zero-root distributions for the same physical state.
By analyzing these distributions together, we obtain an analytical description of the ground-state energy, boundary strings, and bulk excitations of this rank-two model.

The ground-state root configurations consist of bulk strings, boundary strings, and additional roots.
From these configurations, we derive three coupled density equations and obtain the ground-state energy and the surface energy in the thermodynamic limit.
Subtracting the periodic-chain ground-state energy calculated from the Bethe ansatz equations from the open-chain ground-state energy obtained by DMRG yields a finite-size surface energy that approaches the analytical result with $O(1/N)$ corrections.
Beyond determining the ground-state boundary-string positions, the regularity analysis yields a selection rule governing the admissible combinations of boundary strings.
Numerical excited-state calculations identify additional boundary-string configurations allowed by this rule, each associated with a distinct base state.
These base states may also contain bulk excitations required to preserve zero-root symmetry when the number of boundary strings changes.
We further obtain analytical expressions for the boundary-string excitation energies.

For excitations that preserve the boundary-string configuration, we present a zero-root description of three elementary bulk excitations.
Although the boundary conditions are non-diagonal, the excitation energies obtained from the zero-root method agree with the periodic-chain results after accounting for the different normalization \cite{martins}.
We further derive a bulk-excitation selection rule governing their admissible combinations in physical eigenstates, supported by the numerical results.
This bulk-excitation selection rule remains valid even when the reference state already contains holes.

These calculations demonstrate that the DMRG-based extraction of transfer-matrix zeros extends to both the fundamental and fused transfer matrices of the present rank-two model.
Combined with the operator identities, the numerically obtained zeros allow us to calculate the boundary contribution to the energy and the elementary bulk excitation spectrum.
These results provide a basis for extending the approach to other higher-rank integrable chains with off-diagonal boundaries.
A further direction is to investigate finite-size corrections to the energy of the open chain, accounting for both the discreteness of the root distributions and deviations from ideal zero-root configurations.
This would clarify whether, and how, these deviations contribute to the leading $O(1/N)$ correction to the total energy.

\section*{Acknowledgments}

We would like to thank Prof. Y. Wang for valuable discussions and continued encouragement.
We acknowledge financial support from the National Natural Science Foundation of China (Grant Nos. 12675006, 12305005, 12547107, 12434006 and 12247103).

\appendix
\renewcommand{\theequation}{\thesection.\arabic{equation}}
\setcounter{equation}{0}
\section{Fused Reflection Matrix}
\label{app:A}
The elements of the fused reflection matrix $\bar K^{-}(u)$ in Eq.~\eqref{eq:Kfused} are
\begin{equation}
\begin{aligned}
\bar k_{11}&=\frac12\!\left[2(c_2^2-c_1^2)u^2+(3c_2^2+4c_1+c_1^2)u+c_1^2+c_2^2-2\right],\\
\bar k_{12}&=\bar k_{21}=c_2u(2c_1u+c_1-1),\\
\bar k_{13}&=\bar k_{31}=\frac{1}{\sqrt2}u\!\left[2(c_1^2+c_2^2)u+c_1^2-2c_1+c_2^2\right],\\
\bar k_{14}&=\bar k_{41}=-c_2u(2c_1u+c_1-1),\\
\bar k_{15}&=\bar k_{51}=-\frac12u(2u+1)(c_1^2-c_2^2),\\
\bar k_{22}&=\frac12\!\left[4c_1^2u^2+(4c_1^2+2c_2^2)u+c_1^2+c_2^2-2\right],\\
\bar k_{23}&=\bar k_{32}=\sqrt2\,c_2u,\qquad
\bar k_{24}=\bar k_{42}=c_2^2u(2u+1),\\
\bar k_{25}&=\bar k_{52}=c_2u(2c_1u+c_1+1),\\
\bar k_{33}&=\frac12\!\left[2(c_1^2+c_2^2)u+c_1^2+c_2^2-2\right],\\
\bar k_{34}&=\bar k_{43}=-\sqrt2\,c_2u,\\
\bar k_{35}&=\bar k_{53}=-\frac{1}{\sqrt2}u\!\left[2(c_1^2+c_2^2)u+c_1^2+2c_1+c_2^2\right],\\
\bar k_{44}&=\frac12\!\left[4c_1^2u^2+(4c_1^2+2c_2^2)u+c_1^2+c_2^2-2\right],\\
\bar k_{45}&=\bar k_{54}=-c_2u(2c_1u+c_1+1),\\
\bar k_{55}&=\frac12\!\left[2(c_2^2-c_1^2)u^2+(c_1^2-4c_1+3c_2^2)u+c_1^2+c_2^2-2\right].
\end{aligned}
\label{eq:fused-K-elements}
\end{equation}

\section{Numerical Boundary-String Roots}
\label{app:numerical-boundary-roots}
\setcounter{table}{0}
\renewcommand{\thetable}{\thesection.\arabic{table}}

Table~\ref{tab:numerical-boundary-roots} compares the ideal and numerical boundary-root positions in Fig.~\ref{fig:ground-representatives}.
Only positive imaginary parts are listed, rounded to five decimal places; ideal values are calculated from unrounded boundary parameters.

\setlength{\tabcolsep}{3pt}
\renewcommand{\arraystretch}{1.12}
\noindent\makebox[\textwidth][c]{%
\begin{minipage}{0.96\textwidth}
\captionsetup{type=table}
\caption{Boundary-string roots for the configurations in Fig.~\ref{fig:ground-representatives}.
A dash indicates the absence of boundary strings.}
\label{tab:numerical-boundary-roots}
\textbf{(a)}\quad $s_{\rm L}=0.62017,\quad s_{\rm R}=0.29361$
\par\smallskip
\begin{tabular*}{\linewidth}{@{\extracolsep{\fill}}>{\centering\arraybackslash}p{0.13\linewidth}>{\centering\arraybackslash}p{0.37\linewidth}>{\centering\arraybackslash}p{0.19\linewidth}>{\centering\arraybackslash}p{0.19\linewidth}@{}}
\toprule
Root set & Ideal imaginary part & Ideal value & Numerical value \\
\midrule
$\Lambda$ & $s_{\rm R}+3/2$ & $1.79361$ & $1.79361$ \\
 & $s_{\rm L}+3/2$ & $2.12017$ & $2.12105$ \\
\addlinespace
$\bar\Lambda$ & $s_{\rm R}+1$ & $1.29361$ & $1.29362$ \\
 & $s_{\rm L}+1$ & $1.62017$ & $1.62017$ \\
 & $s_{\rm R}+2$ & $2.29361$ & $2.29383$ \\
\bottomrule
\end{tabular*}
\end{minipage}}
\par\medskip
\noindent\makebox[\textwidth][c]{%
\begin{minipage}{0.96\textwidth}
\textbf{(b)}\quad $s_{\rm L}=1.41421,\quad s_{\rm R}=1.16248$
\par\smallskip
\begin{tabular*}{\linewidth}{@{\extracolsep{\fill}}>{\centering\arraybackslash}p{0.13\linewidth}>{\centering\arraybackslash}p{0.37\linewidth}>{\centering\arraybackslash}p{0.19\linewidth}>{\centering\arraybackslash}p{0.19\linewidth}@{}}
\toprule
Root set & Ideal imaginary part & Ideal value & Numerical value \\
\midrule
$\Lambda$ & -- & -- & -- \\
$\bar\Lambda$ & $s_{\rm L}+1$ & $2.41421$ & $2.40128$ \\
 & $s_{\rm R}+1$ & $2.16248$ & $2.16248$ \\
\bottomrule
\end{tabular*}
\end{minipage}}
\par\medskip
\noindent\makebox[\textwidth][c]{%
\begin{minipage}{0.96\textwidth}
\textbf{(c)}\quad $s_{\rm L}=3.92232,\quad s_{\rm R}=1.81071$
\par\smallskip
No boundary strings.
\end{minipage}}
\par\medskip
\noindent\makebox[\textwidth][c]{%
\begin{minipage}{0.96\textwidth}
\textbf{(d)}\quad $s_{\rm L}=0.66519,\quad s_{\rm R}=3.16228$
\par\smallskip
\begin{tabular*}{\linewidth}{@{\extracolsep{\fill}}>{\centering\arraybackslash}p{0.13\linewidth}>{\centering\arraybackslash}p{0.37\linewidth}>{\centering\arraybackslash}p{0.19\linewidth}>{\centering\arraybackslash}p{0.19\linewidth}@{}}
\toprule
Root set & Ideal imaginary part & Ideal value & Numerical value \\
\midrule
$\Lambda$ & $s_{\rm L}+3/2$ & $2.16519$ & $2.16550$ \\
$\bar\Lambda$ & $s_{\rm L}+1$ & $1.66519$ & $1.66519$ \\
\bottomrule
\end{tabular*}
\end{minipage}}
\par

\section{Regularity of the \texorpdfstring{$W$}{W}-Type Root Configuration}
\setcounter{equation}{0}
\label{app:W-center-locking}
To determine the imaginary parts and real centers of the finite roots associated with a $W$-type excitation, we consider
\begin{equation}
\Lambda(u):\quad \pm z_w,
\qquad
\bar\Lambda(u):\quad
\pm\bar z_w\pm i\eta_W,
\qquad
z_w,\bar z_w\in\mathbb{R},\quad \eta_W>0.
\label{eq:W-unlocked-configuration}
\end{equation}
The remaining bulk-string and boundary-root background is kept unchanged.

Subtracting the reference-state equations from the coupled density equations gives the following density responses, with $q=e^{-|\omega|/2}$.
For $\eta_W>1$, they read
\begin{align}
N\delta\tilde\rho_x^{(W)}(\omega)
&=-2\cos(z_w\omega),
\label{eq:W-upper-rho-x}\\
N\delta\tilde\rho_y^{(W)}(\omega)
&=\frac{2\cos(z_w\omega)}q,
\label{eq:W-upper-rho-y}\\
N\delta\tilde\rho_w^{(W)}(\omega)
&=-2q^{2\eta_W-5}\cos(\bar z_w\omega).
\label{eq:W-upper-rho-w}
\end{align}
Regularity requires bounded Fourier responses, whereas Eq.~\eqref{eq:W-upper-rho-y} contains an uncanceled $q^{-1}$ term as $q\to0$.
Thus, the branch $\eta_W>1$ is excluded.

At $\eta_W=1$, adopting the symmetric principal-value prescription for the singular kernels gives
\begin{equation}
N\delta\tilde\rho_x^{(W)}(\omega)
=-2\cos(z_w\omega)
+\frac{q^{-1}+q^3}{1+q^2}\cos(\bar z_w\omega),
\label{eq:W-critical-rho-x}
\end{equation}
\begin{equation}
\begin{aligned}
N\delta\tilde\rho_y^{(W)}(\omega)
={}&\frac{2\cos(z_w\omega)}q
\\
&-\frac{(q^{-2}+q^2)(1+q^2+q^6)}
{(1+q^2)(1+q^6)}\cos(\bar z_w\omega),
\end{aligned}
\label{eq:W-critical-density}
\end{equation}
and
\begin{equation}
N\delta\tilde\rho_w^{(W)}(\omega)
=-\left[q^{-3}+\frac{q^{-1}+q^3}{1+q^6}\right]
\cos(\bar z_w\omega).
\label{eq:W-critical-rho-w}
\end{equation}
Equation~\eqref{eq:W-critical-density} contains a leading $-q^{-2}\cos(\bar z_w\omega)$ term as $q\to0$, which cannot be canceled by the $q^{-1}$ term.
Hence, $\eta_W=1$ is also excluded under the stated prescription.

It remains to consider $0<\eta_W<1$, for which the responses are
\begin{equation}
N\delta\tilde\rho_x^{(W)}(\omega)
=-2\cos(z_w\omega)
+\frac{2(q^{1-2\eta_W}+q^{1+2\eta_W})}{1+q^2}
\cos(\bar z_w\omega),
\label{eq:W-unlocked-rho-x}
\end{equation}
\begin{equation}
\begin{aligned}
N\delta\tilde\rho_y^{(W)}(\omega)
={}&\frac{2\cos(z_w\omega)}q
\\
&-\frac{
2(q^{-2\eta_W}+q^{2\eta_W})(1+q^2+q^6)
}{(1+q^2)(1+q^6)}
\cos(\bar z_w\omega),
\end{aligned}
\label{eq:W-unlocked-rho-y}
\end{equation}
and
\begin{equation}
N\delta\tilde\rho_w^{(W)}(\omega)
=-\frac{2(q^{1-2\eta_W}+q^{1+2\eta_W})}{1+q^6}
\cos(\bar z_w\omega).
\label{eq:W-unlocked-rho-w}
\end{equation}
As $q\to0$, the $y$-density response behaves as
\begin{equation}
N\delta\tilde\rho_y^{(W)}(\omega)
=2q^{-1}\cos(z_w\omega)
-2q^{-2\eta_W}\cos(\bar z_w\omega)
+o(1).
\label{eq:W-locking-asymptotic}
\end{equation}
Regularity requires the exponentially growing terms to cancel, which first fixes $\eta_W=\tfrac12$.
Equation~\eqref{eq:W-unlocked-rho-y} then reduces to
\begin{equation}
N\delta\tilde\rho_y^{(W)}(\omega)
=\frac{2[\cos(z_w\omega)-\cos(\bar z_w\omega)]}{q}
-\frac{2q\cos(\bar z_w\omega)}{1+q^6}.
\end{equation}
Cancellation of the remaining growing term requires $|z_w|=|\bar z_w|$.
Since the roots occur in opposite pairs, their real centers can be chosen nonnegative, giving
\begin{equation}
\eta_W=\tfrac12,
\qquad
\bar z_w=z_w.
\label{eq:W-center-locking}
\end{equation}
Substituting these conditions into Eqs.~\eqref{eq:W-unlocked-rho-x}--\eqref{eq:W-unlocked-rho-w} yields
\begin{align}
N\delta\tilde\rho_x^{(W)}(\omega)
&=0,
\label{eq:W-locked-rho-x}\\
N\delta\tilde\rho_y^{(W)}(\omega)
&=-\frac{2q\cos(z_w\omega)}{1+q^6},
\label{eq:W-locked-rho-y}\\
N\delta\tilde\rho_w^{(W)}(\omega)
&=-\frac{2(1+q^2)\cos(z_w\omega)}{1+q^6}.
\label{eq:W-locked-rho-w}
\end{align}
All three responses are regular and reproduce the $W$-type density variations obtained in the main text.
Thus, within the assumed root configuration, regularity fixes the ideal $W$-type structure as
\begin{equation}
\Lambda(u):\quad \pm z_w,
\qquad
\bar\Lambda(u):\quad
\pm z_w\pm\tfrac{i}{2}.
\label{eq:W-root-configuration-app}
\end{equation}
This result concerns the ideal configuration in the thermodynamic analysis and allows finite-size deviations of the numerical roots.

\section{Symmetry of the Zero-Root Distributions}
\label{app:zero-root-symmetry}
\setcounter{equation}{0}
We take real boundary parameters $c_1,c_2,\tilde c_1,\tilde c_2$ and purely imaginary inhomogeneities, $\theta_j^*=-\theta_j$, including the homogeneous case.
The explicit $R$- and $K$-matrices in Sec.~\ref{sec:integrability} are symmetric and have real polynomial coefficients, giving
\begin{equation}
R_{0j}(u)^\dagger=R_{j0}(u^*),\qquad
K^\pm(u)^\dagger=K^\pm(u^*).
\label{eq:local-hermitian-conjugation}
\end{equation}
The fused matrices obey the same relations, as follows from the explicit $\bar R$-matrix and the symmetric entries of $\bar K^\pm$ in Appendix~\ref{app:A}.
Hermitian conjugation reverses the order of the monodromy factors.
Using $\theta_j^*=-\theta_j$, we obtain
\begin{equation}
\begin{aligned}
T_0(u)^\dagger
&=R_{10}(u^*+\theta_1)\cdots R_{N0}(u^*+\theta_N)
=\hat T_0(u^*),\\
\hat T_0(u)^\dagger&=T_0(u^*).
\end{aligned}
\end{equation}
Substitution into the double-row transfer matrix gives
\begin{equation}
t(u)^\dagger
=\operatorname{tr}_0\!\left[T_0(u^*)K_0^-(u^*)\hat T_0(u^*)K_0^+(u^*)\right]
=t(u^*).
\label{eq:transfer-hermitian-conjugation}
\end{equation}
Here $K_0^+$ acts only in the auxiliary space and can be moved to the front of the partial trace.
Repeating the same steps for the fused transfer matrix gives
\begin{equation}
\bar t(u)^\dagger=\bar t(u^*).
\label{eq:fused-transfer-hermitian-conjugation}
\end{equation}
The commutativity and Hermitian-conjugation relations ensure a common orthonormal eigenbasis for the transfer matrices.
Evaluating these relations in this basis gives
\begin{equation}
\Lambda(u)^*=\Lambda(u^*),\qquad
\bar\Lambda(u)^*=\bar\Lambda(u^*).
\label{eq:eigenvalue-conjugation}
\end{equation}
In terms of the zero-root parameters introduced in Eqs.~\eqref{eq:Lambda-root} and \eqref{eq:Lambda-bar-root}, these identities imply
\begin{equation}
\begin{aligned}
\Lambda\!\left(-\tfrac32+i(-z_j^*)\right)
&=\left[\Lambda\!\left(-\tfrac32+iz_j\right)\right]^*=0,\\
\bar\Lambda\!\left(-\tfrac32+i(-\bar z_j^*)\right)
&=\left[\bar\Lambda\!\left(-\tfrac32+i\bar z_j\right)\right]^*=0.
\end{aligned}
\label{eq:zero-root-reflection}
\end{equation}
Thus, each zero-root parameter $z_j$ is accompanied by $-z_j^*$ in the complete set $\{\pm z_j\}$, and each $\bar z_j$ by $-\bar z_j^*$ in $\{\pm\bar z_j\}$.
These partners have opposite real parts and the same imaginary part, establishing the left--right symmetry of both zero-root distributions about the imaginary axis.

\section{Excitation Configurations and Selection-Rule Verification}
\label{app:numerical-excitation-counts}
\setcounter{table}{0}
\renewcommand{\theHtable}{\theHsection.\arabic{table}}
Table~\ref{tab:numerical-excitation-selection-rule} lists $(n_X,n_Y,n_W)$ for the configurations in Fig.~\ref{fig:excited-state-overlay}.
All configurations satisfy the bulk-excitation selection rule in Eq.~\eqref{eq:excitation-selection-rule}.
\renewcommand{\arraystretch}{1.2}
\begin{longtable}{ccccc}
\caption{Excitation configurations in Fig.~\ref{fig:excited-state-overlay}, specified by $(n_X,n_Y,n_W)$ relative to the common ground state.
The index $\ell=0$ denotes the ground state, and $\ell=1,\ldots,30$ label the excited states.
The last column gives the remainder in the bulk-excitation selection rule; zero indicates that the rule is satisfied.
\label{tab:numerical-excitation-selection-rule}}\\
\toprule
State index $\ell$ & $n_X$ & $n_Y$ & $n_W$ & $(n_X+\tfrac12 n_Y+n_W)\bmod2$ \\
\midrule
\endfirsthead
\multicolumn{5}{c}{Table~\thetable\ (continued)}\\
\toprule
State index $\ell$ & $n_X$ & $n_Y$ & $n_W$ & $(n_X+\tfrac12 n_Y+n_W)\bmod2$ \\
\midrule
\endhead
\midrule
\multicolumn{5}{r}{Continued on next page}\\
\endfoot
\bottomrule
\endlastfoot
$0$ & 0 & 0 & 0 & 0 \\
$1$ & 1 & 2 & 0 & 0 \\
$2$ & 1 & 2 & 0 & 0 \\
$3$ & 1 & 2 & 0 & 0 \\
$4$ & 1 & 2 & 0 & 0 \\
$5$ & 1 & 2 & 0 & 0 \\
$6$ & 1 & 2 & 0 & 0 \\
$7$ & 1 & 2 & 0 & 0 \\
$8$ & 1 & 2 & 0 & 0 \\
$9$ & 1 & 2 & 0 & 0 \\
$10$ & 1 & 2 & 0 & 0 \\
$11$ & 0 & 0 & 2 & 0 \\
$12$ & 1 & 2 & 0 & 0 \\
$13$ & 1 & 2 & 0 & 0 \\
$14$ & 0 & 0 & 2 & 0 \\
$15$ & 0 & 0 & 2 & 0 \\
$16$ & 1 & 2 & 0 & 0 \\
$17$ & 1 & 0 & 1 & 0 \\
$18$ & 1 & 2 & 0 & 0 \\
$19$ & 0 & 0 & 2 & 0 \\
$20$ & 0 & 0 & 2 & 0 \\
$21$ & 0 & 0 & 2 & 0 \\
$22$ & 1 & 2 & 0 & 0 \\
$23$ & 1 & 2 & 0 & 0 \\
$24$ & 1 & 2 & 0 & 0 \\
$25$ & 1 & 2 & 0 & 0 \\
$26$ & 1 & 2 & 0 & 0 \\
$27$ & 0 & 0 & 2 & 0 \\
$28$ & 1 & 0 & 1 & 0 \\
$29$ & 1 & 0 & 1 & 0 \\
$30$ & 1 & 2 & 0 & 0 \\
\end{longtable}

\end{document}